\documentclass[fleqn,usenatbib]{mnras}

\usepackage{newtxtext,newtxmath}

\usepackage[T1]{fontenc}
\usepackage{xcolor}
\usepackage{longtable}

\DeclareRobustCommand{\VAN}[3]{#2}
\let\VANthebibliography\thebibliography
\def\thebibliography{\DeclareRobustCommand{\VAN}[3]{##3}\VANthebibliography}

\DeclareRobustCommand{\DEL}[3]{#2}
\let\DELthebibliography\thebibliography
\def\thebibliography{\DeclareRobustCommand{\DEL}[3]{##3}\DELthebibliography}

\usepackage{graphicx}	
\usepackage{amsmath}	
\usepackage{multirow}

\definecolor{ao(english)}{rgb}{0.0, 0.5, 0.0}

\title[SAMI: Drivers of stellar-gas misalignments]{The SAMI Galaxy Survey: Physical drivers of stellar-gas kinematic misalignments in the nearby Universe}

\author[A. Ristea et al.]{A. Ristea,$^{1,2}$\thanks{E-mail: andrei.ristea@icrar.org (AR)}
L. Cortese,$^{1,2}$
A. Fraser-McKelvie,$^{1,2}$
S. Brough,$^{2,3}$
J. J. Bryant,$^{2,4,5}$
B. Catinella,$^{1,2}$ 
\newauthor
S. M. Croom,$^{2,4}$
B. Groves,$^{1,2}$
S. N. Richards,$^{4}$
J. van de Sande$^{2,4}$
J. Bland-Hawthorn,$^{2,4,5}$
\newauthor
M. S. Owers,$^{2,6,7}$
and J. S. Lawrence$^{8}$
\\
$^{1}$International Centre for Radio Astronomy Research, The University of Western Australia, 35 Stirling Highway, Crawley WA 6009, Australia\\
$^{2}$ARC Centre of Excellence for All Sky Astrophysics in 3 Dimensions (ASTRO 3D) \\
$^{3}$School of Physics, University of New South Wales, NSW 2052, Australia \\
$^{4}$Sydney Institute for Astronomy (SIfA), School of Physics, The University of Sydney, NSW 2006, Australia \\
$^{5}$Australian Astronomical Optics, AAO-USydney, School of Physics, University of Sydney, NSW 2006, Australia \\
$^{6}$School of Mathematical and Physical Sciences, Macquarie University, NSW 2109, Australia \\
$^{7}$Astronomy, Astrophysics and Astrophotonics Research Centre, Macquarie University, Sydney, NSW 2109, Australia \\
$^{8}$Australian Astronomical Optics, Macquarie University, 105 Delhi Rd, North Ryde, NSW 2113, Australia
}

\date{Accepted XXX. Received YYY; in original form ZZZ}

\pubyear{2022}

\begin{document}
\label{firstpage}
\pagerange{\pageref{firstpage}--\pageref{lastpage}}
\maketitle

\begin{abstract}
Misalignments between the rotation axis of stars and gas are an indication of external processes shaping galaxies throughout their evolution. Using observations of 3068 galaxies from the SAMI Galaxy Survey, we compute global kinematic position angles for 1445 objects with reliable kinematics and identify 169 (12\%) galaxies which show stellar-gas misalignments. Kinematically decoupled features are more prevalent in early-type/passive galaxies compared to late-type/star-forming systems. Star formation is the main source of gas ionisation in only 22\% of misaligned galaxies; 17\% are Seyfert objects, while 61\% show Low-Ionisation Nuclear Emission-line Region features. We identify the most probable physical cause of the kinematic decoupling and find that, while accretion-driven cases are dominant, for up to 8\% of our sample, the misalignment may be tracing outflowing gas. When considering only misalignments driven by accretion, the acquired gas is feeding active star formation in only $\sim$1/4 of cases. As a population, misaligned galaxies have higher S\'ersic indices and lower stellar spin \& specific star formation rates than appropriately matched samples of aligned systems. These results suggest that both morphology and star formation/gas content are significantly correlated with the prevalence and timescales of misalignments. Specifically, torques on misaligned gas discs are smaller for more centrally concentrated galaxies, while the newly accreted gas feels lower viscous drag forces in more gas-poor objects. Marginal evidence of star formation not being correlated with misalignment likelihood for late-type galaxies suggests that such morphologies in the nearby Universe might be the result of preferentially aligned accretion at higher redshifts.
\end{abstract}

\begin{keywords}
galaxies: general -- galaxies: evolution -- galaxies: statistics -- galaxies: kinematics and dynamics 
\end{keywords}



\section{Introduction}
\label{sec:intro}
In the Lambda Cold Dark Matter ($\Lambda$CDM) paradigm, galaxies form from perturbations in the primordial density field. Baryons collapse into the gravitational potential of dark matter halos and gain angular momentum from interactions with the background tidal field, as predicted by tidal torque theory (TTT) \citep{Hoyle1951,peebles1969,Doroshkevich1970}. If the angular momentum is weakly conserved during this process, a stable rotating gaseous disc will take shape, from which stars will then form (\citealt{FallEfstathiou1980}; \citealt{Mo1998}). The newly formed stars will inherit the kinematic properties of the gas, which will be depleted through star formation.

While depletion timescales are typically of the order of a few Gyrs, a significant fraction of galaxies in the nearby Universe are observed to be still forming stars (\citealt*{KennicuttTamblynCongdon1994}), implying a replenishment of the gas supply to maintain star formation (\citealt*{PutmanPeekJong2012}; \citealt{SanchezAlmeida2014}).

Galaxies can acquire new gas through both internal and external processes. Internal mechanisms that replenish the gas supply refer to stellar mass loss through e.g. supernovae explosions or stellar winds, and in such cases, the newly acquired gas inherits the dynamical properties of the stars. External processes include cold gas accretion from filaments (e.g. \citealt{Keres2005,Chung2012}), accretion of hot gas from a galaxy's outer halo (e.g. \citealt{CdpLagos2015}) or clumpy accretion from gas-rich galaxy interactions/mergers (e.g. \citealt{DiTeodoro2014,Barreraballesteros2015}). Recycled gas may also appear to have an external origin when re-accreted in a galactic fountain \citep{HouckBregman1990}. The newly acquired gas from these processes can have an angular momentum offset from that of the stars by any angle, potentially resulting in an observed stellar-gas kinematic misalignment. Consequently, such kinematic features provide evidence for a possible external origin of the gas, making galaxies with kinematically decoupled stellar-gas rotation the ideal sample for studying the processes of gas accretion and settling, and their impact on galaxy evolution. 

Gas accreted in a misaligned configuration will feel a torque from the stellar disc, causing it to settle into a stable state over a given timescale. Such a process was simulated by \cite{Freekevdv2015} for a $10^{11}\ \rm{M_{\odot}}$ early-type galaxy, who found that misaligned configurations remain in place while new gas is being accreted since the angular momentum of the incoming material is dominant. After accretion stops, the torque exerted by the stellar disc causes the gas to settle into a stable, co-rotating state. Furthermore, the counter-rotating configuration between stars and gas (i.e rotation in the same plane, in opposite directions, corresponding to a position angle offset of $180^{\rm{o}}$) was also found to be stable in observational studies (e.g. \citealt{Bryant2019}; \citealt{Duckworth}). Notably, \cite{OsmanBekki2017} found that gas in this state can remain stable until it is consumed by star formation. Such a configuration can result from accretion of a gas reservoir in a retrograde orbit, with the accreted gas dominating the in-situ component in terms of mass (\citealt{Kannappan_Fabricant2001}; \citealt{Bassett2017}; \citealt{Bryant2019}). 

The time that a misaligned gas disc spends in an unstable state before settling can vary greatly, depending on the stellar mass distribution of a galaxy. Intuitively, more centrally concentrated distributions will result in smaller torques on misaligned gas discs, thus increasing the lifetime of the decoupled configuration. Such results have been ubiquitously reported in observational studies (e.g. \citealt{Davis2015,Bryant2019,Duckworth,Zhou2022}), who found that misalignments are more frequent in galaxies with early-type morphologies (i.e. with more centrally concentrated stellar mass distributions). The correlation between morphology and misalignments has also been extensively analysed in simulation-based works set in the $\Lambda$CDM framework, with similar results (e.g. \citealt{CdpLagos2015}; \citealt{Khim2020I}; \citealt{Casanueva2022}).
    
In a misaligned accretion event, the incoming gas will interact with the in-situ component, resulting in viscous drag forces between the two gas reservoirs that will cause a decrease in angular momentum, speeding up the settling process. Notably, this process is expected to affect misaligned gas at any angle, including in a counter-rotating configuration \citep{Davis2011}. Circumstantial evidence of this phenomenon was previously reported in studies based on the Sloan Digital Sky Survey IV (SDSS-IV) - Mapping Nearby Galaxies at Apache Point Observatory (MaNGA) Galaxy Survey (\citealt{Bundy2015}), in the form of misaligned galaxies having either lower angular momentum or rotational to dispersion support than their aligned counterparts, for both stars and gas \citep{Chen2016,Jin2016,Duckworth,Xu,Zhou2022}. The direct effect of cold gas in driving misalignment timescales has been previously addressed in simulation-based works, where the computation of such quantities is straightforward, with \cite{CdpLagos2015} and \cite{Khim2020I} finding that galaxies with lower gas fractions are more likely to display kinematically decoupled configurations. 

If the cold gas fraction is associated with the prevalence of misalignments, one would expect such features to occur less often in more star-forming galaxies. While the correlation between morphology and misalignment likelihood is relatively well established in observational work, the contribution of star formation remains uncertain. Since the two properties are expected to be interconnected, a study of the \textit{independent} effects of morphology and star formation/gas content on the likelihood of misalignment development is required in order to establish whether one property is dominant over the other. 

Furthermore, if newly accreted misaligned gas is forming stars, it will be depleted within a given timescale equal to the gas mass divided by the rate of star formation. \cite{Davis2015} proposed that the absence of a significant number of counter-rotating objects in a sample of galaxies from the $\rm{ATLAS^{3D}}$ project \citep{Capellari2011} implies that the accreted gas is depleted by star formation before it can settle into this configuration. While observational evidence of a peak in the misalignment angle distribution around the counter-rotating region has been since reported (\citealt{Bryant2019}), the extent to which newly acquired gas traced
by kinematic misalignments in the  nearby Universe is feeding new star formation is still not well established.

Finally, while samples of galaxies displaying stellar-gas kinematic misalignments are archetypal when studying the phenomenon of gas accretion and settling, it is important to note that such features can also be caused by gas being expelled from the galaxy in an Active Galactic Nucleus (AGN)-driven galactic outflow (e.g. \citealt{Dumas2007}; \citealt{RongxinLuo2019L}) or ram-pressure stripping events (e.g. \citealt{Bryant2019}). Typically, in previous observational-based studies of misaligned galaxies, the assumption was made that gas accretion (from either filaments, outer halo, or galaxy interactions/mergers) is the exclusive physical cause of the feature \citep{Davis2011,Davis2015,Jin2016,Duckworth}, while \cite{Bryant2019} also considered the possibility of gas stripping causing kinematic decouplings.  \cite{BarreraBallesteros2014,Barreraballesteros2015} focused on the effect of galaxy interactions/mergers on producing kinematic misalignments between stars and gas. Simulation-based work in the $\rm{\Lambda}$CDM framework, such as \cite{CdpLagos2015} and \cite{Casanueva2022} similarly consider external gas accretion (and internal processes, in the case of the latter study) to be the cause(s) of stellar-gas misalignments. \cite{Khim2020II} also take into account group environmental effects and interactions with the brightest group galaxy as potentially resulting in a kinematic decoupling between stars and gas. 
When selecting a sample of misaligned galaxies with externally acquired gas for a statistical analysis, it is important to consider the possibility of gas expulsion being the cause of the kinematic decoupling, in case such features contribute to observed numbers by a non-negligible fraction.

In this paper, we make use of the final data release from the Sydney-AAO Multi-object Integral-field spectrograph (SAMI) Galaxy Survey (\citealt{Bryant2015}; \citealt{Croom2021}) and identify galaxies exhibiting misalignments between the stellar and ionised gas rotation. Using a combination of the spectral classification of misaligned galaxies, their position angle offset measurement and optical morphology, we determine the most probable physical cause of the misalignment. This technique provides an estimate of the contribution of gas inflows and outflows to observed misalignment numbers. Finally, we study the independent correlations between morphology, star formation, and  the prevalence and timescales of stellar-gas kinematically decoupled configurations.

This paper is structured as follows: Section 2 presents the SAMI Galaxy Survey data, the quality cuts implemented, kinematic position angle computation and final sample selection process; Section 3 describes our results regarding the distributions of PA offsets for different galaxy populations, spectral properties of misaligned galaxies, the most probable physical processes causing them and the influence of morphology and star formation on misalignment timescale and prevalence; Section 4 discusses our results in the context of gas accretion and settling while Section 5 summarises our findings and provides concluding remarks.
Throughout this paper, we assume a flat $\rm{\Lambda}$CDM concordance cosmology: $H_0=70\ \rm{km\ s^{-1}\ Mpc^{-1}}$, $\Omega_0=0.3$, $\Omega_{\Lambda}=0.7$.


\section{Data and Methodology}
\label{methods}

The data reduction process employed for SAMI observations is described in \cite{Sharp2015}, \cite{Allen2015} and \cite{Green2018}. In summary, the \texttt{2DFDR} pipeline is used (\citealt*{CroomSaundersHeald2004}; \citealt{SharpBirchall2010}), complemented by further corrections for telluric absorption features and atmospheric dispersion, made using secondary standard stars. The resulting row-stacked spectra are then combined into an IFS data cube of 50 $\rm{\times}$ 50 spaxels of width $0.5^{\prime\prime}$ $\rm{spaxel^{-1}}$.

In this work, we employ the two-moment Gaussian line-of-sight velocity distribution  stellar \citep{vds2017} and H$\alpha$ \citep{Ho2016} kinematic maps, including the rotational velocity and velocity dispersion ($\sigma$) velocity maps. The stellar signal-to-noise ratio (hereafter $S/N_{\rm{stars}}$) was calculated as the median across the entire blue wavelength range, from the flux and variance spectra in each spaxel. The H$\alpha$ signal-to-noise ratio (hereafter $S/N_{\rm{H\alpha}}$) was computed as the ratio of the emission line intensity to its associated error, as given by the linear least-squares fitting process. The average seeing full-width at half maximum (FWHM) of the point-spread function (PSF) for the SAMI DR3 sample is $\sim2^{\prime\prime}$.   

\begin{figure*}
	\centering
	\includegraphics[width=\linewidth]{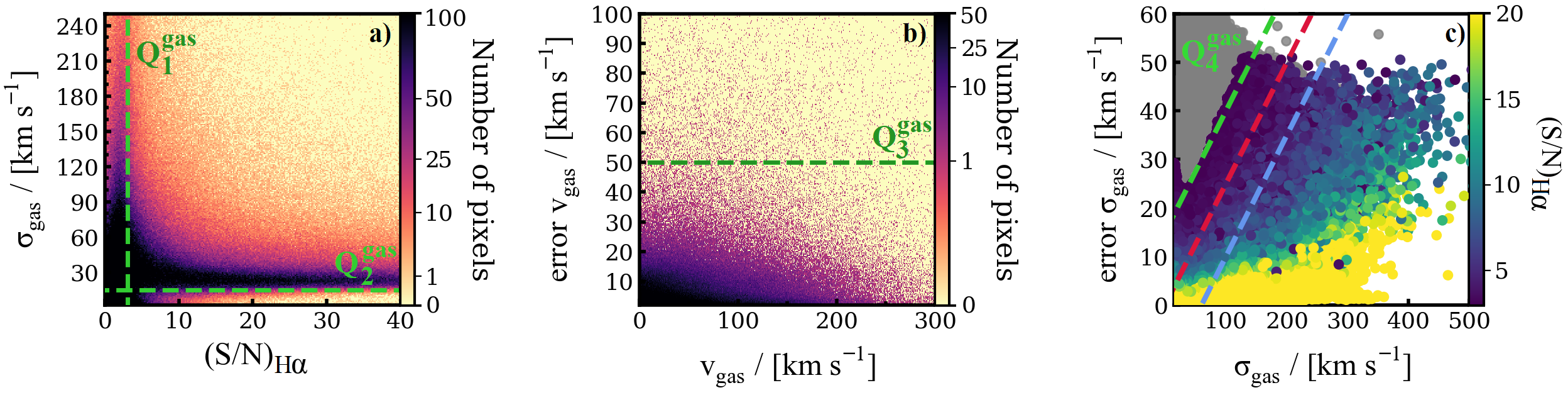}
    \caption{Illustration of quality cuts applied to spaxels in gas velocity maps: \textbf{(a)} 2D histogram showing the H${\alpha}$ ionised gas velocity dispersion ($\sigma_{\rm{gas}}$) of spaxels in SAMI DR3 as a function of their H$\alpha$ S/N. The vertical and horizontal dashed green lines show the $Q_{\rm{1}}^{\rm{gas}}$ and $Q_{\rm{2}}^{\rm{gas}}$ quality cuts, respectively. \textbf{(b)} 2D histogram showing the error in the ionised gas rotational velocity ($v_{\rm{gas}}$) vs the respective velocity value for spaxels that have passed the $Q_{\rm{1}}^{\rm{gas}}$ and $Q_{\rm{2}}^{\rm{gas}}$ quality cuts. The horizontal dashed green line shows the $Q_{\rm{3}}^{\rm{gas}}$ selection criterion. \textbf{(c)} Scatter plot showing the error in $\sigma_{\rm{gas}}$ vs $\sigma_{\rm{gas}}$. Spaxels passing $Q_{\rm{1}}^{\rm{gas}}$ - $Q_{\rm{3}}^{\rm{gas}}$ are shown, color-coded according to their $S/N_{\rm{H\alpha}}$, while those with $(S/N_{\rm{H\alpha}})<3$ are shown in grey. The dashed lines correspond to prospective quality cuts: \textcolor{ao(english)}{(green)} err$(\sigma) = 0.25\times \sigma + 15\ \rm{km\ s^{-1}}$ ($Q_{\rm{4}}^{\rm{gas}}$); \textcolor{red}{(red)} err$(\sigma) = 0.25\times \sigma$; \textcolor{blue}{(blue)} err$(\sigma) = 0.25\times \sigma - 15\ \rm{km\ s^{-1}}$. }
    \label{QCs_plots}
\end{figure*}

\subsection{Quality cuts and parent sample selection}
\label{sec:QC} 

The degree of certainty to which we can determine global kinematic position angles (PAs) is influenced by the quality of the resolved stellar and gas kinematic fits. In turn, the quality of these fits depends on a number of factors, including the S/N of spectra, the brightness of the H$\alpha$ emission line and how close the velocity dispersion measure is to the instrumental resolution. Similarly to \cite{vds2017}, who analysed the stellar kinematics of galaxies in SAMI DR2, we apply quality cuts to the stellar velocity maps based on the S/N, rotational velocity ($v_{\rm{stars}}$) and dispersion ($\sigma_{\rm{stars}}$) uncertainties.

We initially exclude all stellar velocity map spaxels with $S/N_{\rm{stars}} < 5$ in the stellar continuum (hereafter $\boldsymbol{ {Q_{1}^{\star}} }$). This selection is applied in order to eliminate the spaxels where high random uncertainties result in unreliable velocity measurements which may influence our PA estimates. Following this cut, we implement the same selection criteria for stellar velocity maps as \cite{vds2017} (with one exception as mentioned) and exclude all spaxels with: ($\boldsymbol{Q_{2}^{\star}}$) $\sigma_{\rm{stars}} < 35\ \rm{km\ s^{-1}}$ (see \citealt{Fogarty2015}); ($\boldsymbol{Q_{3}^{\star}}$) err$\ (v_{\rm{stars}}) > 50\ \rm{km\ s^{-1}}$ (instead of a more stringent cut at $30\ \rm{km\ s^{-1}}$ since the computation of global kinematic PAs is less sensitive than the high-order stellar kinematics moments computed by \citealt{vds2017}); ($\boldsymbol{Q_{4}^{\star}}$) err$\ (\sigma_{\rm{stars}})<0.1 \times \sigma_{\rm{stars}}\ +\ 25\ \rm{km\ s^{-1}}$. The four selection criteria for stellar velocity maps ($Q_{1}^{\star}-Q_{4}^{\star}$) are met by 41.3\% of SAMI DR3 spaxels in stellar velocity maps.

Building on the work of \cite{vds2017}, we implement similar quality cuts for the ionised gas rotational velocity maps in SAMI DR3. Figure \ref{QCs_plots} (a) shows the H$\alpha$ velocity dispersion versus $(S/N)_{\rm{H\alpha}}$ for all 2159409 spaxels in the gas velocity maps in SAMI DR3. We exclude all gas velocity map spaxels with $S/N_{\rm{H\alpha}} < 3$, as shown by the vertical dashed line in Figure \ref{QCs_plots} (a) ($\boldsymbol{Q_{1}^{\rm{gas}}}$). To minimise the effect of systematic errors in spaxel spectra where the dispersion is close to the instrumental resolution, we further exclude all spaxels with $\sigma_{\rm{gas}} < 15\ \rm{km\ s^{-1}}$, as indicated by the horizontal dashed line in Figure \ref{QCs_plots} (a) ($\boldsymbol{Q_{2}^{\rm{gas}}}$). This cut excludes the spaxels in the high density region between $3 \lesssim S/N_{\rm{H\alpha}} \lesssim 5$ and below the $Q_{2}^{\rm{gas}}$ line in Figure \ref{QCs_plots} (a), without the need of a stricter cut in  $S/N_{\rm{H\alpha}}$ which would bias our data toward low $\sigma_{\rm{gas}}$ values. We note that, while lower than the SAMI spectral resolution, a higher threshold in $\sigma_{\rm{gas}}$ for $Q_{2}^{\rm{gas}}$ was found to exclude a significant number of high-$S/N_{\rm{H\alpha}}$ spaxels. After implementing $Q_{1}^{\rm{gas}}$ and $Q_{2}^{\rm{gas}}$, the majority of the remaining spaxels have velocity uncertainties below $50\ \rm{km\ s^{-1}}$. As shown in Figure \ref{QCs_plots} (b), the remaining spaxels with err$ (v_{\rm{gas}}) > 50\ \rm{km\ s^{-1}}$ tend to have relatively high fractional uncertainties in the gas rotational velocity and are excluded from the gas velocity maps ($\boldsymbol{Q_{3}^{\rm{gas}}}$ - dashed line in Figure \ref{QCs_plots} b).

Next, similarly to $Q_{4}^{\star}$, we consider a number of quality cuts based on the ionised gas velocity dispersion and its uncertainty. These cuts are shown in Figure \ref{QCs_plots} (c) by the green (err$\ \sigma_{\rm{gas}} < 0.25\ \times \ \sigma_{\rm{gas}} + 15\ \rm{km\ s^{-1}}$), red (err$\ \sigma_{\rm{gas}} < 0.25\ \times \ \sigma_{\rm{gas}}$) and blue (err$\ \sigma_{\rm{gas}} < 0.25\ \times \ \sigma_{\rm{gas}} - 15\ \rm{km\ s^{-1}}$) dashed lines respectively, such that all the spaxels above each line would be excluded (spaxels shown are those remaining after the application of $ Q_{1}^{\rm{gas}}-Q_{3}^{\rm{gas}}$). 

We find that the quality cut corresponding to the blue line tends to severely bias the resulting spaxel sample toward large velocity dispersions. For the red and green lines, the differences in the mean $\sigma_{\rm{gas}}$ of the remaining spaxels, after applying the cuts, are $\Delta\sigma_{\rm{ mean,red}}^{\rm{after-before}} = 1.79\ \rm{km\ s^{-1}}$  and  $\Delta\sigma_{\rm{mean,green}}^{\rm{after-before}} = 0.11\ \rm{km\ s^{-1}}$. We therefore adopt the quality cut shown by the green dashed line in Figure \ref{QCs_plots} (c) ($\boldsymbol{Q_{4}^{\rm{gas}}}$). This selection criterion excludes all the remaining spaxels with low ($\sim20-25\ \rm{km\ s^{-1}}$) gas velocity dispersions and high associated uncertainties ($\sim20-30\ \rm{km\ s^{-1}}$) in the dark blue region above the $Q_{4}^{\rm{gas}}$ line, without biasing our resulting spaxel sample towards large $\sigma_{\rm{gas}}$ values. The four selection criteria applied to gas velocity maps ($ Q_{1}^{\rm{gas}}-Q_{4}^{\rm{gas}} $) are met by 44.8\% of the initial number of spaxels in the gas velocity maps. 
Note that these need not be the same spaxels that pass the stellar kinematics quality cuts, and in fact only 25.2\% (18.7\%) of spaxels in the original stellar (gas) velocity maps meet both sets of quality criteria (i.e. are common for both kinematic maps).

Following the implementation of the quality cuts, we exclude from any further analysis all 1324 galaxies where the number of spaxels in either the stellar or gas velocity maps are below 50 and 100, respectively. This criterion takes into account the fact that our QCs have the effect of removing a relatively larger number of spaxels from stellar velocity maps than from gas, while also ensuring that the spatial scale probed by the kinematics is, in all cases, larger than $\sim 2\ \times\ $FWHM.  We further inspect all galaxies where the number of spaxels in the stellar/gas velocity maps is between 50-100/100-200, and exclude a further 299 objects where the global kinematic trend in either component is uncertain due to low data quality. This results in a final \textit{parent sample} of 1445 galaxies (1141 from the GAMA regions and 304 from the cluster sample) for which we compute both stellar and gas kinematic PAs, as detailed in Section \ref{sec:PAs_measurements}. 


\subsection{Kinematic position angles}
\label{sec:PAs_measurements}

In this work, we compute global kinematic PAs using both the Kinemetry-based method described in \cite{Krajnovic2006Kinemetry}, and a technique which implements the Radon transform (\citealt{Stark2018Radon}). A brief description of each technique, and its associated advantages and shortcomings, is presented in this section. We also discuss the procedure used to decide which of the two methods is more appropriate for tracing the global rotation of each galaxy. For both methods, kinematic PAs are measured from North = 0, counter-clockwise to the receding side of the velocity map, in the range $0^{\rm{o}}-360^{\rm{o}}$.

Kinemetry is a parametric method of modelling the line-of-sight velocity distribution of a galaxy which also allows for the computation of global PAs, and is described in \cite{Krajnovic2006Kinemetry}. The latter is implemented through the \texttt{FIT$\_$KINEMATIC$\_$PA}\footnote{\url{https://www-astro.physics.ox.ac.uk/~mxc/software/}} routine developed by \cite{Cappellari2007} and \cite{Krajnovic2011} (hereafter, referred to as \textit{Kinemetry}). This method works with the underlying assumption that the velocity field is bi-antisymmetric. However, physical processes such as past mergers, tidal interactions, gas accretion, stripping, etc., or non-axisymmetric morphological features such as bars, can cause disturbances in the rotation fields of both stars and gas. In some cases of galaxies in this work, such perturbations cause deviations from bi-antisymmetry, while also maintaining a global velocity gradient which allows for the measurement of a PA. 

To alleviate any potential shortcomings associated with Kinemetry, we also compute kinematic PAs using the Radon Transform method introduced by \cite{Stark2018Radon}. Briefly, this is a non-parametric method of computing global kinematic PAs which works by integrating the difference between the velocity of each spaxel along a given line passing through the kinematic centre, and the mean velocity along that line. The angle between the direction that minimises the line integral, and the y-axis, is then equal to the kinematic PA. The integration is performed within a set boundary called the Radon aperture. Larger values of the Radon aperture tend to reduce random errors by being less sensitive to small-scale perturbations (e.g. turbulent motions) in the velocity fields. On the other hand, as the Radon aperture increases, variations in the PA get smoothed out. As such, the choice of the integration interval is a matter of balancing out these two effects. 

Similarly to \cite{Stark2018Radon}, we use a value for the Radon aperture equal to $R_{\rm{e}} \times (b/a)$, where $R_{\rm{e}}$ is the circularised half-light radius and $b/a$ is the minor-to-major axial ratio (see Section \ref{structural_param}). This choice ensures that the integration is performed over similar physical scales, projected onto the face of each galactic disc. In the cases where the Radon aperture is smaller than the spatial resolution of the SAMI DR3 observations, we set it to be equal to the FWHM of the PSF. 
While the Radon method makes no assumptions about the underlying rotational structure of a galaxy, it tends to be quite sensitive to relatively small numbers of spaxels with unphysical velocity values, which have been omitted by our quality cuts.  

For the parent sample passing our quality cuts (1445 galaxies - see Section \ref{sec:QC}), we compute kinematic PAs using both the Kinemetric ($PA_{\rm{Kin}}$) and Radon ($PA_{\rm{Radon}}$) methods. The resulting distribution of $PA_{\rm{Kin}}-PA_{\rm{Radon}}$ has a median of $0.0^{\rm{o}}$ for both stellar and gas rotation, confirming that there is no systematic offset between the methods. The $16^{\rm{th}}$ and $84^{\rm{th}}$ percentiles of $PA_{\rm{Kin}}-PA_{\rm{Radon}}$ are $(-10^{\rm{o}},13^{\rm{o}})_{\rm{stars}}$ and $(-13^{\rm{o}},13^{\rm{o}})_{\rm{gas}}$ for the two components, respectively. We visually inspect the velocity fields where the  disagreement between the two methods is larger than the standard deviation of the $PA_{\rm{Kin}}-PA_{\rm{Radon}}$ distribution for stars ($23^{\rm{o}}$) and gas ($22^{\rm{o}}$), and decide on the method which best determines the global kinematic PA. This is the case for 181 (13\%) and 144 (10\%) out of 1445 stellar and gas velocity fields, respectively. All such cases have been quality-selected and thus have ordered rotation in both components, with velocity gradients clearly visible in their kinematic maps, constrained within  1 $\sigma$ ($22^{\rm{o}}-23^{\rm{o}}$). The discrepancies in PA estimates for these objects are due to a clear systematic failure of one of the methods, with the other PA measurement providing a reliable description of the velocity gradient orientation. Out of these, Radon kinematic PAs were found by visual inspection to be the best choice for 42 (23\%) and 43 (30\%) stellar and gas rotation fields, respectively, showing an overall higher reliability of Kinemetry. As a result, for the cases where the difference in PAs between the two methods is within one standard deviation, we choose the kinematic PA given by Kinemetry. 

Stellar and gas velocity maps of galaxies in our parent sample extend, on average, out to 1.2 and 1.7 effective radii. As a consequence, the analysis presented in the next section identifies definitive cases of stellar-gas  global misalignments and is not sensitive to potential decouplings between the two components at larger radii.

\subsection{Misaligned sample definition }
\label{sec:misaligned}

The projected misalignment angle between the stellar and gas kinematic PAs is computed as $\Delta PA_{\rm{stars-gas}} = PA_{\rm{stars}} - PA_{\rm{gas}}$, with $PA_{\rm{stars}}$ and $PA_{\rm{gas}}$ being the stellar and gas kinematic PAs, calculated as described in Section \ref{sec:PAs_measurements}.

Similarly to previous studies (e.g. \citealt{Davis2011}; \citealt{CdpLagos2015}; \citealt{Jin2016}; \citealt{Bryant2019}; \citealt{Casanueva2022}; \citealt{Xu}; \citealt{Zhou2022}), we define a galaxy as being kinematically misaligned if $|\Delta PA_{\rm{stars-gas}}| \geq 30^{\rm{o}}$. While this distinction is somewhat arbitrary, we note that such a threshold accounts for the PA measurement errors for galaxies in our sample. As mentioned in Section \ref{sec:PAs_measurements}, the $16^{\rm{th}}$ and $84^{\rm{th}}$ percentiles of $PA_{\rm{Kin}}-PA_{\rm{Radon}}$ distribution are $(-10^{\rm{o}},13^{\rm{o}})_{\rm{stars}}$ and $(-13^{\rm{o}},13^{\rm{o}})_{\rm{gas}}$, respectively. In addition to this uncertainty, our measured PAs are projected values and the 3-dimensional misalignments will be further affected by orientation. Notably, \cite{Bryant2019} and \cite{Duckworth} tested the effect of increasing the misalignment threshold from $30^{\rm{o}}$ to $40^{\rm{o}}$ and reported that this does not affect the main conclusions of their works significantly.

Furthermore, 21 galaxies in our parent sample have unconstrained stellar and/or gas kinematic PAs. These cases are either galaxies where the respective kinematics have been affected by recent mergers, accretion events or cluster processes, or slow rotators in the stellar component (or alternatively, a combination of these situations). The physical processes that generate such kinematic features are the same as those that cause misalignments, where both the stellar and gas components are rotating in a disc and global PAs can be measured. As such, we include these galaxies (hereafter referred to as \textit{kinematically unconstrained}) in our misaligned sample, with the caveat that these systems do not have an associated $\Delta PA_{\rm{stars-gas}}$ value. The resultant misaligned sample consists of 169 galaxies (148 having a misalignment angle measurement and 21 being kinematically unconstrained). 

\subsection{Ancillary data}
\label{sec:ancillary_data} 

\subsubsection{Stellar mass and star formation rate}
\label{sec:Ms_SFR}

Given that the SAMI survey sample is a combination of galaxies extracted from the GAMA survey footprint and eight galaxy clusters, there are no homogeneous estimates of global star formation rates (SFR) for the entire DR3 sample. It has been shown that using SFR estimates based on the H$\alpha$ flux for SAMI galaxies can be significantly affected by contamination from AGN emission and aperture corrections \citep{Medling2018}.
To improve the situation with respect to previous SAMI studies, here we have taken advantage of three different SFR estimates to obtain the best catalogue of homogenized SFRs for the entire SAMI sample. We stress that our goal here is not to have very accurate (i.e. < 0.1 dex) SFRs, but to be able to determine in which area of the $M_{\rm{\star}}$ - SFR plane our misaligned sample lies. As discussed below, the homogenization procedure provides us with SFRs having statistical uncertainties of the order of 0.25 dex or less. This value is smaller than the typical scatter in the star-forming main sequence (SFMS) and, as such, more than enough for our goals.  
 
Following on from \cite{FM2021}, we start from the Galaxy Evolution Explorer (\textit{GALEX}) SDSS Wide-field Infrared Survey Explorer (\textit{WISE}) Legacy Catalog version 2 (GSWLC-2; \citealt{Salim2016}; \citealt*{Salim2018}) which provides SFRs and stellar masses obtained via the spectral energy distribution (SED)-fitting Code Investigating GALaxy Emission (CIGALE) (\citealt*{Burgarella2005}; \citealt{Noll2009}; \citealt{Boquien2019}) for 1891 SAMI galaxies in both the GAMA and cluster fields. We use these galaxies to derive an empirical recipe to rescale the stellar masses of SAMI galaxies presented in \cite{Bryant2015} to the GSWLC-2 stellar masses. For galaxies not in GSWLC-2, we determine stellar masses by rescaling the values presented in \cite{Bryant2015}, using the derived recipe (the median and standard deviation of the rescaling are both $\sim$0.1 dex). This method provides us with an homogeneous set of stellar mass values for all SAMI DR3, with a typical scatter of $\sim$0.14 dex. To compute SFRs, we proceed as follows:

For GAMA galaxies with no SFRs in GSWLC-2, we take advantage of the recent compilation by \cite{Bellstedt2020} and \cite{Driver2022}, presenting SFRs obtained using the SED-fitting code ProSPECT (\citealt{Robotham2020prospect}). For galaxies in common, the two indicators agree with a scatter of $\sim$ 0.24 dex, but show a systematic offset of $\sim$ 0.1 dex which we add to the ProSPECT values to correct for this effect. This discrepancy is not surprising, due to the different photometry and SED-fitting techniques used. In addition, it is worth noting that the two fitting codes deal very differently with passive galaxies. In GSWLC-2, most passive galaxies do not show specific star formation rates (sSFRs) below $10^{-13}\ \rm{yr^{-1}}$, while ProSPECT values do reach several orders of magnitudes lower sSFRs. As such, to be consistent with GSWLC, we force galaxies with sSFRs in ProSPECT below $10^{-13}\ \rm{yr^{-1}}$ to this threshold value. The combination of GSWLC-2 and ProSPECT SFRs provides us with SFRs for 2548 galaxies out of the 3068 in the SAMI DR3 sample.
For an additional 81 galaxies detected by both \textit{GALEX} and \textit{WISE} but not included in the catalogues listed above, we follow \cite{Janowiecki2017} and determine global SFRs by combining NUV photometry with \textit{WISE} W3 fluxes. The comparison of these SFR estimates with those from GSWLC-2 shows a scatter of 0.2 dex and a systematic offset of 0.2 dex that we correct for by adding it to our derived SFRs.
In summary, our technique allows us to determine homogenized SFRs for 2629 galaxies, corresponding to 86\% of the whole SAMI DR3 sample. Out of the 1445 galaxies in our parent sample, 1369 (95\%) have SFR measurements available. For the misaligned sample, 163/169 galaxies (96\%) have SFR values computed.

\subsubsection{Structural parameters }
\label{structural_param}

Effective radii and ellipticities for galaxies in the SAMI DR3 sample were determined using the Multi-Gaussian expansion (MGE, \citealt*{Emsellem1994}) technique, implemented in the code from \cite{Capellari2002} (see \citealt{Deugenio2021}). Throughout this paper, we make use of the circularised effective radius ($R_{\rm{e}}$) and the minor-to-major axial ratio ($b/a$). S\'ersic indices (\citealt{Sersic1963}) are used in this work to quantify the shape of a galaxy's stellar mass distribution, in order to study its contribution to the prevalence of the kinematic misalignment phenomenon. These were computed from a single-component S\'ersic fit to the \textit{r}-band photometry by \cite{Lee2012} for GAMA galaxies, and \cite{Owers2019} for the cluster sample. Out of 1445 galaxies in our parent sample, 1405 (97\%) have reliable S\'ersic index measurements. For the misaligned sample, this is the case for 165/169 galaxies (98\%).

\subsubsection{Stellar spin parameter $\lambda_{Re}$}
In this study, we use a proxy for the  spin parameter (inclination-corrected) within one effective radius ($\lambda_{Re}$) to quantify the ratio of rotational to dispersion motion in each galaxy. It was previously shown (see \citealt{Cappellari_review} for a review) that $\lambda_{Re}$ is a physically-motivated proxy for galaxy structure, informative in particular in the case of massive systems where other indicators such as the S\'ersic index and bulge-to-total ratio fail to discriminate between rotationally- and dispersion-dominated systems.
To this end, we use $\lambda_{Re}$ in this work as an indicator of morphology, complementary to the single-component S\'ersic index.
Following the method described in \cite{Emsellem2007} and \cite{Emsellem2011}, we compute the spin parameter $\lambda_{Re}$ as the flux-weighted ratio of ordered to disordered motion within a galaxy:

\begin{equation}
    \lambda_{Re}=\frac{\langle R|V|\rangle}{\langle R\ \sqrt[]{V^2 + \sigma^2} \rangle} = \frac{\sum_{i=0}^{N_{spx}} F_iR_i|V_i|}{\sum_{i=0}^{N_{spx}} F_iR_i\ \sqrt{V_{i}^{2} + \sigma_{i}^{2}}} ,
	\label{eq:lambda_re}
\end{equation}

\noindent where $F_i$,  $V_i$ and  $\sigma_i$ are the flux, stellar rotational velocity and stellar velocity dispersion of the $i^{th}$ spaxel respectively, $N_{\rm{spx}}$ is the number of spaxels within 1 $R_{\rm{e}}$ and $\rm{\langle \rangle}$ denotes a luminosity count weighting. $R_i$ is defined as the semi-major axis of the ellipse on which the $i^{th}$ spaxel lies (\citealt{Cortese2016}). The quality cuts applied to velocity maps and criteria required for $\lambda_{Re}$ values to be considered reliable are described in \cite{vds2017}. Out of the 1445 galaxies in the parent sample, 1149 (80\%) pass the quality criteria for computing $\lambda_{Re}$ estimates. This is the case for 123/169 (72\%) galaxies in the misaligned sample.

The computed $\lambda_{Re}$ values are influenced by both the seeing conditions (FWHM of the PSF), and by a galaxy's inclination angle with respect to the line-of-sight (\citealt{Cappellari2016}, \citealt{Graham2018}). To account for seeing effects, the correction of \cite{Harborne2020} was applied, optimised for the SAMI Galaxy survey as described by \cite{jvds2021}. When correcting for inclination, we use the method of \cite{Emsellem2011}, in the manner implemented by \cite{delMoralCastro2020}. This method is dependent on a galaxy's edge-on axis ratio ($q_{\rm{0}}$), for which we use the values adopted by \cite{Cortese2016} depending on whether a galaxy has a clear disc component or not (the SAMI visual morphological classification technique is outlined in \citealt{Cortese2016}). The inclination correction also takes into account a galaxy's anisotropy parameter in the meridional plane ($\beta_{\rm{z}}$), for which we use a value of $\beta_{\rm{z}}=0.3$ \citep{Santucci2022}.


\begin{table*}
\centering
\caption{Number of galaxies and percentage of the SAMI DR3, parent and misaligned samples that have $n_{\rm{S\Acute{e}rsic}}$, $\lambda_{Re}$ and SFR measurements available, as well as the split of each respective sample into star forming, non star-forming, late- and early-type galaxies.}\label{tab1}
\begin{tabular*}{\textwidth}{@{\extracolsep{\fill}}ccccccc|cc} 
\hline
\multirow{2}{*}{Sample}      & \multicolumn{3}{c}{Parameter completeness} & \multirow{2}{*}{Total}  & \multicolumn{4}{c}{Split between sub-samples}  \\
                             & $n_{\rm{S\Acute{e}rsic}}$  & $\lambda_{Re}$     & SFR                         & \multicolumn{1}{l}{} & SF &  Non SF & Late-type & Early-type                                        \\ 
\hline
SAMI DR3                     & 2963 (97\%) & 1824 (59\%) & 2629 (86\%) & 3068  &  1619 (62\%) & 1010 (38\%) & 1760 (59\%) & 1203 (41\%)                                                      \\
Parent sample (PAs computed) & 1399 (97\%) & 1149 (80\%) & 1369 (95\%)                      & 1445 & 982 (72\%) & 387 (28\%) & 947 (68\%) & 452 (32\%)                                                           \\Misaligned sample            & 165 (98\%)  & 123 (73\%)  & 163 (96\%)                         & 169 &  43 (25\%) & 120 (75\%) & 28 (17\%) & 137 (83\%)                                                         \\
\hline
\end{tabular*}
\label{Table1}
\end{table*}

\subsection{Spectral classification from emission line properties}
\label{sec:spectral_class_method}

In this work, we use the SAMI aperture spectra to identify the main source of gas ionisation for our kinematically decoupled galaxies. We  classify our misaligned sample spectroscopically, based on the main gas ionisation mechanism, into star-forming (SF), Low-Ionisation Nuclear Emission-line Region (LINER; \citealt{Heckman1980}) and Seyfert objects. To this end, we use the  [\ion{O}{iii}]$\lambda$5007/H$\beta$, [\ion{N}{ii}]$\lambda$6583/H$\alpha$ and [\ion{S}{ii}]$\lambda \lambda$6717,31/H$\alpha$ (hereafter [\ion{O}{iii}]/H$\beta$, [\ion{N}{ii}]/H$\alpha$ and [\ion{S}{ii}]/H$\alpha$) emission line ratios from integrated spectra within 1 $R_{\rm{e}}$. Individual emission line intensities from aperture spectra were computed using the \texttt{LZIFU} fitting software (see \citealt{Ho2016}). Measurements within 1 $R_{\rm{e}}$ were not available for two galaxies in our misaligned sample, cases in which we used the spectra integrated within a $4^{\prime\prime}$ aperture. For one of these objects (CATID 31740), the half-light radius could not be measured reliably, while for the other (CATID 23265), the $4^{\prime\prime}$ aperture is a factor of 1.2 times
larger than 1 $R_{\rm{e}}$. The latter object has an H$\beta$ intensity that is too weak within 1 $R_{\rm{e}}$, and hence we use the $4^{\prime\prime}$ aperture measurements for all emission lines, for consistency. We only consider an emission line to be detected if the S/N within the given integrated spectrum is > 5. Where this is not the case, we estimate the respective emission line intensity as being equal to I = 5 $\times$ err(I), which represents an upper limit on its value. In our misaligned sample, the H$\beta$, [\ion{S}{ii}], [\ion{N}{ii}] and [\ion{O}{iii}] emission lines are not detected up to S/N = 5 in 43, 18, 1 and 1 case(s), respectively.

\section{Results}

Throughout this section, we refer to the SAMI DR3, parent (Section \ref{sec:QC}) and misaligned (Section \ref{sec:misaligned}) samples of galaxies in relation to a number of physical parameters outlined in Section \ref{sec:ancillary_data}. The completeness to which each of these parameters are measured for each respective sample, given the quality of the data, is summarised in Table \ref{tab1}. Stellar mass measurements are available for all galaxies in SAMI DR3. 

\begin{figure*}

	\centering
	\includegraphics[width=\linewidth]{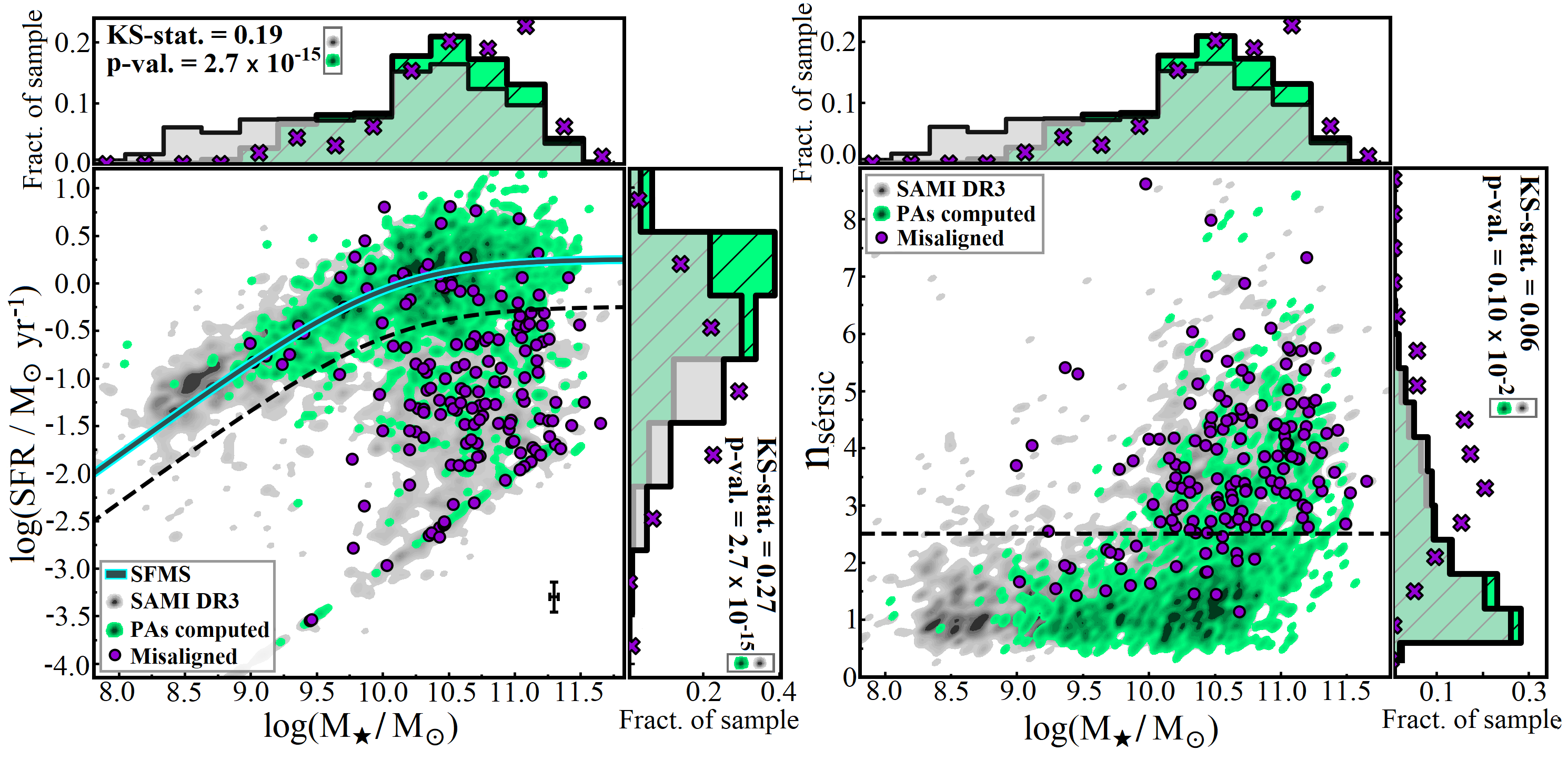}
    \caption{The $M_{\star}$ - SFR (\textbf{left}) and $M_{\star}-n_{\rm{S\Acute{e}rsic}}$ (\textbf{right}) planes for the misaligned (purple), parent (green) and full SAMI DR3 (grey) samples. The teal-cyan solid line shows the star-forming main sequence, as computed by \protect\cite{FM2021}. The black dashed lines are placed, in the left and right panels, at 0.5 dex below the star-forming main sequence and at $n_{\rm{S\Acute{e}rsic}}$ = 2.5, and are used in this work to separate the star-forming/non-star-forming and late/early-type populations, respectively. The distributions of SFR, S\'ersic index and stellar mass for each sample (normalised to the total sample size) are shown adjacently. The KS-statistics and p-values displayed are computed from a Kolmogorov-Smirnoff test comparing the respective parameter distributions of the SAMI DR3 sample (grey), and the parent sample (green), as indicated by the markers placed next to the values. The black error bars in the left plot show median uncertainties in $M_{\star}$ and SFR for the parent sample.}
    \label{fig:Mstar_SFR_nsersic_plots}
\end{figure*}

\subsection{Physical properties of parent and misaligned samples}
\label{sec:physical_properties}

The parent and misaligned samples are shown on the $M_{\star}$ - SFR and $M_{\star}-n_{\rm{S\Acute{e}rsic}}$ planes in Figure \ref{fig:Mstar_SFR_nsersic_plots}, together with the whole SAMI DR3 sample. 
We note that our quality cuts introduce a bias against low $M_{\star}$ as shown by a statistical comparison of the parent and SAMI DR3 samples using a Kolmogorov-Smirnoff (KS) test (p-value = 2.7 $\times \ 10^{-15}$). This bias is introduced because low-$M_{\star}$ ($\lesssim 10^9 \rm{M_{\odot}}$) galaxies tend to have lower S/N in the stellar continuum \citep{Bryant2019}. There is also a bias introduced by our quality cuts towards galaxies with higher SFRs (p-value = 2.7 $\times\ 10^{-15}$) due to the fact that we require a sufficiently high S/N in the H$\alpha$ line in order to measure PAs, and this emission line arises mostly from ionized gas associated with star-forming regions. 

We use a threshold at 0.5 dex below the SFMS (dashed line in Figure \ref{fig:Mstar_SFR_nsersic_plots} - left) to separate the galaxies in the two samples into star forming (SF - above the threshold) and non star forming (non-SF - below the threshold). 

Furthermore, we classify the misaligned and parent sample galaxies in this work based on their morphology, into early- and late-types. This separation is done based on a cut in S\'ersic index at $n_{\rm{S\Acute{e}rsic}}$ = 2.5 (dashed line in Figure \ref{fig:Mstar_SFR_nsersic_plots} - right), with galaxies above this threshold being classified as early-types and those below, late-types. 
The results of these separations into SF/non-SF and late/early-type sub-samples are summarised in Table \ref{tab1}. Overall, we find that a significantly higher fraction of misaligned galaxies are early-types/non SF than late-types/SF, while the opposite is the case for our parent sample (a result contributed to by our QC-introduced biases discussed above). 


\begin{figure*}
	
	\centering
	\includegraphics[width=\linewidth]{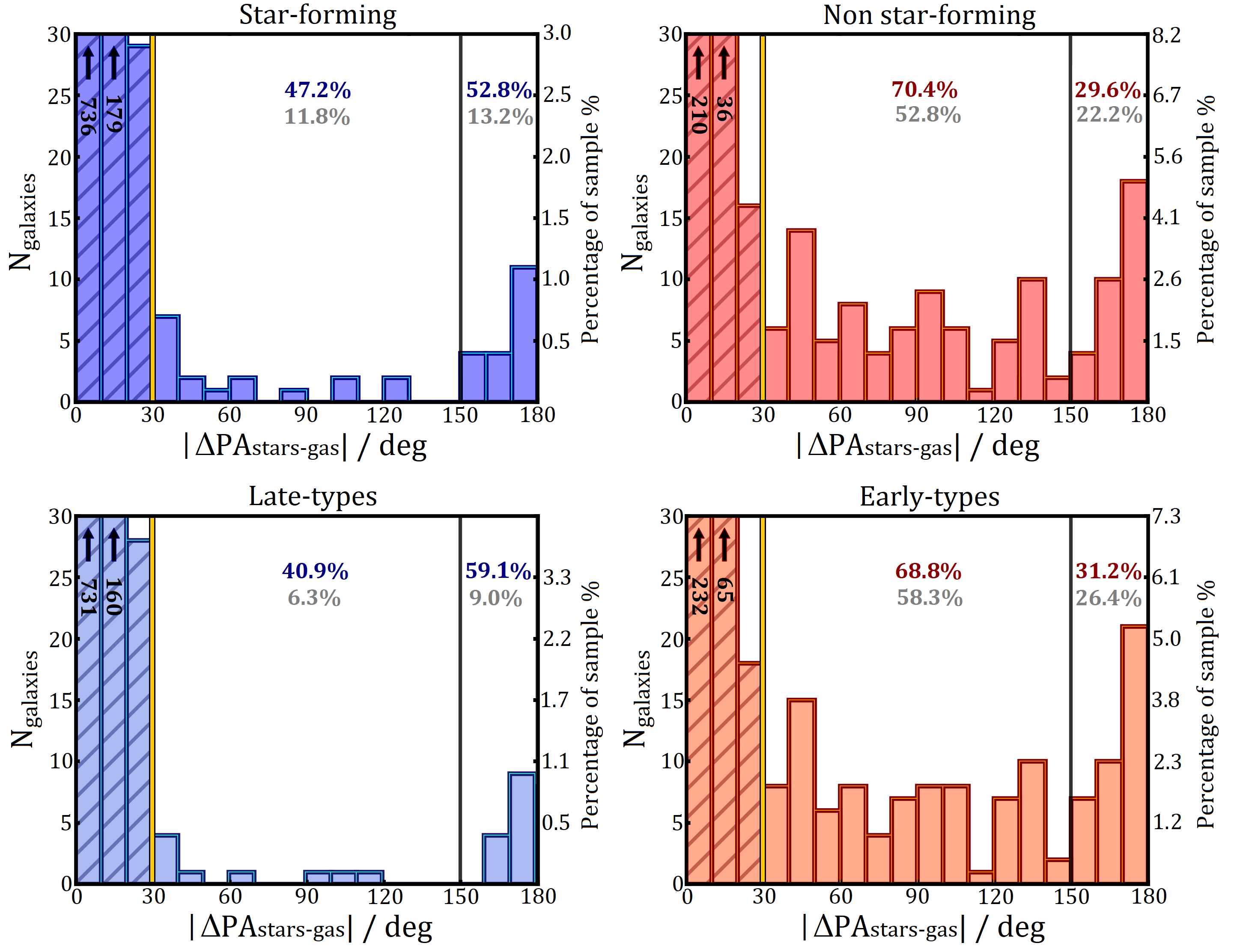}
    \caption{Distributions of the absolute values of misalignment angles between stellar and gas rotation, $|\Delta PA_{\rm{stars-gas}}|$ (Section \ref{sec:misaligned}) for our parent sample, split into star-forming (\textbf{top left}) and non star-forming (\textbf{top right}). The bottom row shows the distributions for the parent sample, split into late-types (\textbf{bottom left}) and early-types (\textbf{bottom right}). The vertical axis on the right side of each panel shows the corresponding fraction of the respective sub-sample. The hatched bins show the galaxies with aligned stellar-gas rotation, separated from the misaligned sample by the vertical yellow line. The vertical black line shows the separation between the two misalignment categories: unstable configuration $[30^{\rm{o}}-150^{\rm{o}}]$ and counter-rotating $(150^{\rm{o}}-180^{\rm{o}}]$. The percentages in red/blue show the number of misaligned galaxies in each respective angular range relative to the total number of misaligned objects in that sub-sample (i.e. SF, non-SF, late- and early-type), while those shown in grey are computed relative to the entire misaligned sample.    }
    \label{fig:dPA_plots}
\end{figure*}

\subsection{Misalignment angle distribution}
\label{sec:dpa_distribution}

A total of 12 \% of galaxies in our parent sample show global stellar-gas kinematic misalignments. 
Figure \ref{fig:dPA_plots} shows the distributions of $|\Delta PA_{\rm{stars-gas}}|$ for parent sample galaxies with misalignment angle measurements available (Section \ref{sec:misaligned}), split into SF and non-SF (top row) sub-samples, as well as early- and late-types (bottom row). We first note that the fraction of SF parent sample galaxies which are misaligned is 4\%, while this number goes up to 31\% for non-SF objects. When splitting our sample by morphology, we find that only 3\% of late-type galaxies are misaligned, while for early-type morphologies, this fraction is 30\%. 

In this work, we distinguish between two different types of misaligned galaxies which are expected to trace different stages in the process of gas settling, or alternatively, different physical processes causing them (Section \ref{sec:misalign_cause}):

\begin{itemize}
    \item \textbf{Unstable configuration misalignments} $[30^{\rm{o}}-150^{\rm{o}}]$ : Galaxies which are tracing misaligned gas still in the process of stabilising into the co- or counter-rotating configurations, or alternatively, gas being expelled from the galaxy in an outflow (see Section \ref{sec:misalign_cause});
    \item \textbf{Counter-rotating misalignments} $(150^{\rm{o}}-180^{\rm{o}}]$ : Galaxies where gas has either been accreted in a counter-rotating configuration, or alternatively, galaxies where the misaligned gas disc had a different kinematic offset angle initially, and has now precessed into the stable counter-rotating state (within PA measurement uncertainties). 
\end{itemize}

We decide not to isolate a third class of polar rings/close-to-perpendicular misalignments as none of the distributions in Figure \ref{fig:dPA_plots} show deviations from being uniform between $60^{\rm{o}}-120^{\rm{o}}$ at a significant level (with p-values in the range 0.29-0.97 when comparing with a uniform distribution using a KS-test). Conversely, we find a counter-rotating peak that is always statistically significant (with p-values below 0.9$\ \times\ 10^{-3}$ when comparing with a uniform distribution using a KS-test), motivating the choice to separate clearly unstable from stable configurations. 

The counter-rotating peak for the whole misaligned sample has a size equal to 4\% of the co-rotating peak. For both the non-SF and early-type misaligned sub-samples, this fraction increases to $\sim$12\%, while for the SF and early-type populations, we find that the counter-rotating galaxies represent only 2\% and 1\% of the kinematically aligned peak, respectively. These ratios are summarised in Table \ref{tab:TableA1} (Appendix \ref{A0}), together with their associated errors.

The fraction of galaxies in our parent sample that display a kinematic decoupling still in the process of migrating to a co- or counter-rotating state (i.e. in an unstable configuration, $30^{\rm{o}}\leq |\Delta PA_{\rm{stars-gas}}|\leq 150^{\rm{o}} $) is 7\%. For the early-type and non-SF sub-samples individually (Figure \ref{fig:dPA_plots} - right column), this percentage increases to 19\% and 20\%, respectively. Conversely, for the SF and late-type populations (Figure \ref{fig:dPA_plots} - left column), the fractions of galaxies in an unstable misaligned configurations are only 2\% and 1\%, respectively. 

A breakdown of the percentages of parent sample galaxies (and their uncertainties) found in each misaligned category, for the different populations in Figure \ref{fig:dPA_plots}, is shown in Table \ref{tab:TableA1} (Appendix \ref{A0}).

We note that there is no visible difference in the misalignment angle distribution shapes between the SF and late-type populations, within the misaligned angular range $[30^{\rm{o}}-180^{\rm{o}}]$, or equivalently between the non-SF and early-types. A comparison of the $|\Delta PA_{\rm{stars-gas}}|$ distributions using a KS-test reveals that they are statistically similar for the SF and late-type populations (p-value = 0.66); this is also the case for the non-SF and early-types (p-value = 0.99). The statistical similarity is also shown by the percentages of misaligned galaxies in each category (unstable and counter-rotating configurations) being consistent between the mentioned sub-sample distributions (as displayed on Figure \ref{fig:dPA_plots}). However, this is not a result of the two populations being the same as we note that only 11/22 (50\%) of misaligned late-type galaxies with PAs computed are classified as SF, while 96/122 (79\%) misaligned early-type galaxies are also in the non-SF sub-sample.

We find a clear difference in the shapes of the misalignment angle distributions between SF and non-SF populations (top row of Figure \ref{fig:dPA_plots}), as well as between the late- and early-type sub-samples (bottom row of Figure \ref{fig:dPA_plots}), within $[30^{\rm{o}}-180^{\rm{o}}]$. A statistical comparison shows dissimilarities between the $|\Delta PA_{\rm{stars-gas}}|$ distributions of the SF and non-SF populations (p-value = 0.06), which is also the case when comparing the late- and early-types (p-value = 0.02). Notably, the statistical difference is driven by the fact that a higher fraction of kinematically decoupled objects are counter-rotating in the SF ($ \sim$ 53\%) and late-type ($\sim$59\%) misaligned sub-samples than for the non-SF and early-types ($\sim$30\% and $\sim $31\%), respectively.

In summary, our $|\Delta PA_{\rm{stars-gas}}|$ distribution is consistent with two separate classes of misaligned objects: stable (counter-rotating) and unstable. The exact shapes of the distributions appear to be sensitive to both morphology and SFR. We will further investigate these correlations in Sections \ref{sec:fractions} and \ref{sec:distributions_results}; the next step in our analysis is to investigate the spectral properties of misaligned galaxies and determine what the physical origin of kinematically decoupled features in our sample is.

\begin{figure*}
	\centering
	\includegraphics[width=\linewidth]{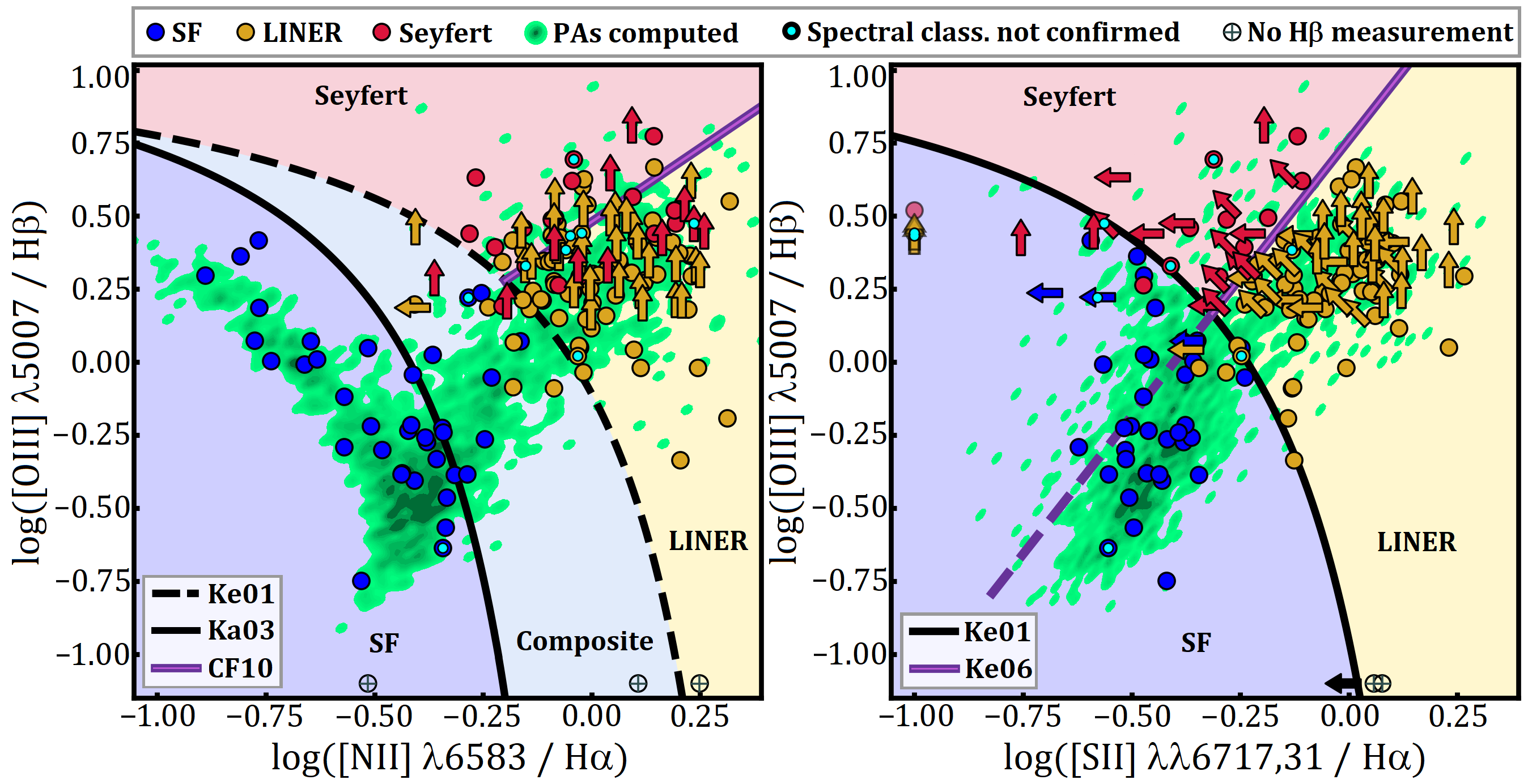}
    \caption{[\ion{N}{ii}]-BPT (\textbf{left}) and [\ion{S}{ii}]-BPT (\textbf{right}) for our misaligned and parent samples. Circles show misaligned galaxies for which all lines have been detected at $S/N > 5$ (or higher) while arrows indicate upper/lower limits of the emission line ratios.  The demarcations used to separate various classes are shown in the [\ion{N}{ii}]-BPT by the the solid black (Ka03), solid purple (CF10) and dashed black (Ke01) lines, and in the [\ion{S}{ii}]-BPT by the black (Ke01) and purple (Ke06) lines. Points are color-coded according to their final classification. Galaxies where no H$\beta$ intensity measurement is available are placed at -1.1 on the vertical axis (circles with crosses), while those with no [\ion{S}{ii}] detected are placed at -1.1 on the horizontal axis. The parent sample galaxies are shown in green. }
    \label{BPT_plots}
\end{figure*}

\subsection{Spectral classes of misaligned galaxies}
\label{sec:spectral_results}

Figure \ref{BPT_plots} shows the Baldwin, Phillips \&  Terlevich (BPT) diagrams (\citealt*{BPT1981}) plotting the [\ion{O}{iii}]/H$\beta$ emission line ratio versus [\ion{N}{ii}]/H$\alpha$ (left panel, hereafter [\ion{N}{ii}]-BPT) and [\ion{S}{ii}]/H$\alpha$ (right panel, hereafter [\ion{S}{ii}]-BPT) for our misaligned sample. We use a combination of the two emission line diagrams for our spectral classification. We employ the semi-empirical prediction of \citeauthor{Kauffmann2003} (\citeyear{Kauffmann2003}; hereafter Ka03) for the extreme starburst limit (black solid line on the [\ion{N}{ii}]-BPT) to identify the SF-dominated objects. The Ka03 limit represents a revision of the \citeauthor{Kewley2001} (\citeyear{Kewley2001}; hereafter Ke01) upper starburst threshold and we consider the region between the two as containing \textit{composite} objects where ionization is due to both stars and AGN. 

\citeauthor{Kewley2006} (\citeyear{Kewley2006}; hereafter Ke06) identified a split of the right wing in the [\ion{N}{ii}]-BPT, into Seyfert and LINER branches, which is clearly more visible on the [\ion{S}{ii}]-BPT. This separation occurs because the [\ion{N}{ii}]/H$\alpha$ line ratio saturates at high metallicities where it is no longer a linear function of the nebular metallicity (\citealt{KewleyDopita2002}; \citealt*{Denicolo2002})
To distinguish between the Seyfert and LINER galaxies in our sample, we employ the (semi-empirical) separation determined by Ke06 and classify all galaxies above the Ke01 extreme starburst limit on the [\ion{S}{ii}]-BPT (black line), and above/below the Ke06 delimitation (purple line) as Seyfert/LINER galaxies. The galaxies which do not fall in any of the above categories are classified using a combination of the two BPTs, as follows:

\begin{itemize}
\item galaxies in the SF region on the [\ion{S}{ii}]-BPT and the composite region on the [\ion{N}{ii}]-BPT are classified as \textbf{SF};
\item galaxies in the SF region below the Ke06 line on the [\ion{S}{ii}]-BPT, and in the LINER region on the [\ion{N}{ii}]-BPT are classified  as \textbf{LINERS}. The LINER-Seyfert delimitation on the [\ion{N}{ii}]-BPT was made using the parametrisation of \citeauthor{RCidFernandes2010} (\citeyear{RCidFernandes2010}; CF10);  
\item galaxies in the SF region and above the Ke06 line on the [\ion{S}{ii}]-BPT, and in either the LINER or Seyfert region on the [\ion{N}{ii}]-BPT are classified  as \textbf{Seyferts}.  
\end{itemize}

We note that three galaxies in our misaligned sample are classified solely based on their position on the [\ion{N}{ii}]-BPT due to not having integrated [\ion{S}{ii}] measurements in any aperture. A further three objects have unconstrained [\ion{O}{iii}]/H$\beta$ emission line ratios and are therefore labelled as having an uncertain spectral classification.

To confirm that the spectral properties within 1 $R_{\rm{e}}$ are representative for the whole spatial scale probed by the kinematics, we analyse resolved [\ion{N}{ii}]/H$\alpha$ and [\ion{S}{ii}]/H$\alpha$ emission-line ratio maps for our misaligned sample, matched to the sizes of H$\alpha$ rotational velocity maps, after the application of QCs (including the exclusion of pixels where any of the emission lines has S/N<3). Our spectral classifications are not confirmed for 9 galaxies in the sample (2 SF, 4 LINERs and 3 Seyfert objects) and we label these galaxies as having an \textit{unconfirmed spectral classification} (cyan dots in Figure \ref{BPT_plots}). We note that the two misaligned objects for which integrated spectra within $4^{\prime \prime}$ apertures (instead of 1 $R_{e}$) were used (Section \ref{sec:spectral_class_method}) both have their spectral classification confirmed by this check.

The result of our spectral classification is displayed in Figure \ref{BPT_plots}. These findings are also summarised in Table \ref{tab:tablemaster}, where uncertainties in the reported numbers and fractions reflect the galaxies with unconfirmed spectral classifications from the analysis of their resolved emission line ratio maps presented above. 
Our results show that energetic feedback from newly formed stars is not the dominant mechanism for hydrogen ionization in misaligned galaxies, with only 22 $\%$ of such galaxies being spectrally classified as SF (blue circles in Figure \ref{BPT_plots}).
The majority of misaligned galaxies in our sample are in the non-star-forming region of the parameter space. According to our classification scheme, 61 $\%$ of kinematically decoupled galaxies have a dominant LINER component, while the remaining 17 $\%$ are classified as Seyferts.

Finally, we note that 31/37 (84 $\%$) of misaligned galaxies where ionization is dominated by star formation are also classified as being SF based on their position with respect to the SFMS. From the remaining 129 galaxies with an AGN component, only 12 (9\%) are classified as being SF from the $M_{\star}$ - SFR plane. For the analysis presented in Sections \ref{sec:fractions} and \ref{sec:distributions_results}, the separation between SF and non-SF objects is made based on their position with respect to the SFMS as we are interested in how a galaxy's cold gas content (for which our best available proxy is sSFR) is associated with the prevalence and timescales of misalignment.

\subsection{Physical causes of stellar-gas kinematic misalignments}
\label{sec:misalign_cause}

\begin{figure*}
	\centering
	\includegraphics[width=\linewidth]{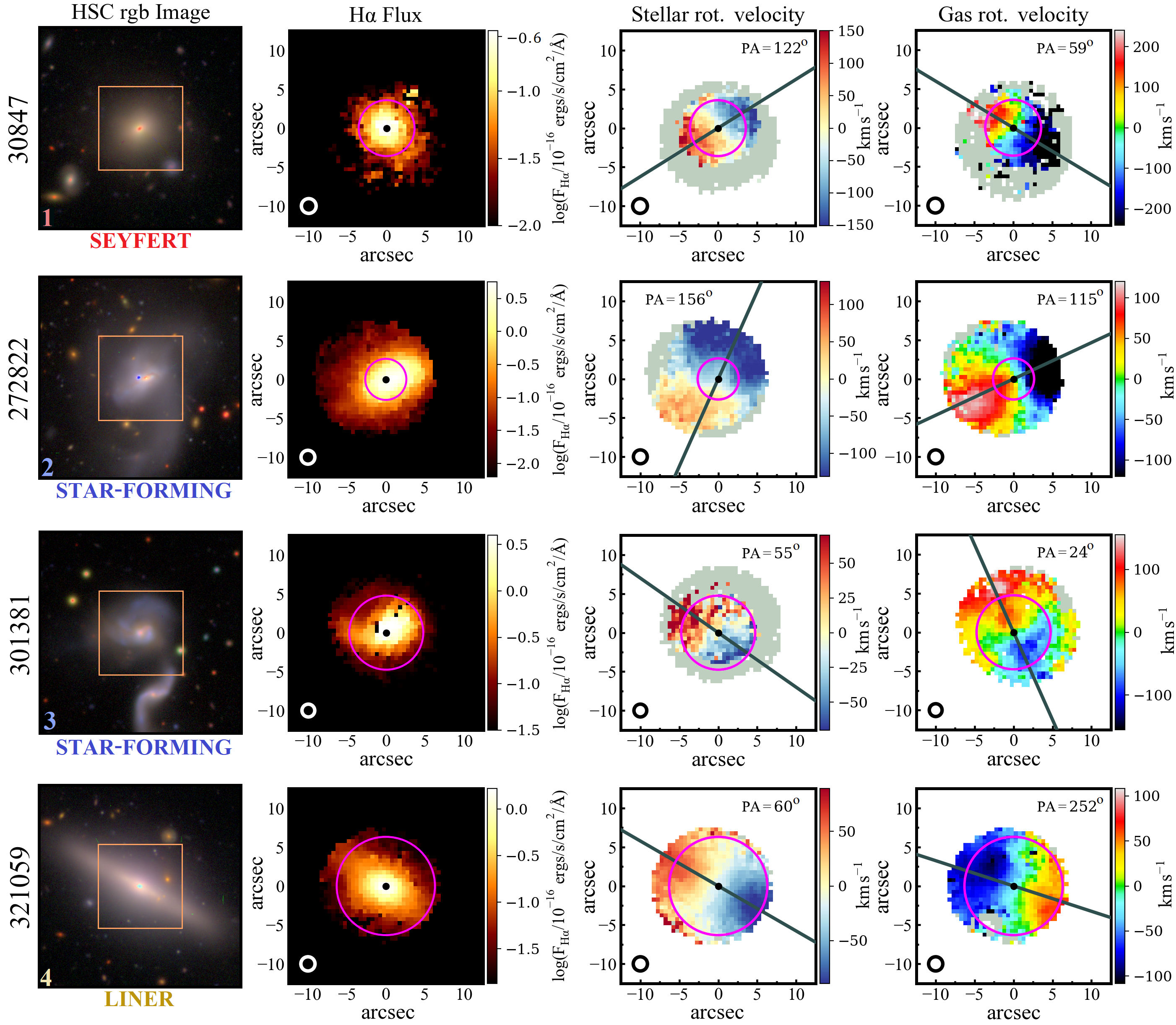}
    \caption{Example misaligned galaxies in SAMI DR3, showcasing each of the dominant physical causes of the kinematic decoupling, classified as described in Section \ref{sec:misalign_cause}: \textbf{(1) CATID 30847}: gas outflow (Seyfert object with $|\Delta PA_{\rm{stars-gas}}|\ =\ 63^{\rm{o}}$); \textbf{(2) CATID 272822}: recent merger (galaxy in a post-merger stage); \textbf{(3) CATID 301381}: tidal interaction / close pair; \textbf{(4) CATID 321059}: gas accretion from other sources, i.e. filaments, outer halo or a past merger (this galaxy does not fall in any of the above three categories). The columns show, from left to right: HSC rgb image of the galaxy; SAMI H$\alpha$ flux map; stellar rotational velocity map; ionised gas rotational velocity map. The number in the bottom-left corner of the HSC image corresponds to its position on the $M_{\star}$ - SFR plane in Figure \ref{fig:Misalignment_cause_plot}. The spectral class (Section \ref{sec:spectral_results}) of each galaxy is highlighted under the HSC image.  The white/black circle in the bottom left corner of the H$\alpha$ intensity and velocity maps shows the FWHM of the PSF, while the magenta circles show the extent of the half-light radius ($R_{\rm{e}}$) around the optical centre. The kinematic PA of each stellar and gas velocity map is shown by the dark green line, with the angle's value (measured counterclockwise from North = 0) displayed on the plot. The background spaxels shown in light green on the velocity maps have been exclude by the quality cuts.    }
    \label{fig:Galaxy_maps_examples}
\end{figure*}

Before investigating the physical properties and potential origins of our misaligned sample, it is important to first determine the origin of the misaligned gas. To this end, we employ the PA offset measurements (Section \ref{sec:PAs_measurements}) and information on the gas ionisation source (Section \ref{sec:spectral_results}) previously derived, together with the optical imaging at our disposal.
We make use of optical Hyper Suprime-Cam (HSC) rgb images for misaligned galaxies in the GAMA fields, and a combination of SDSS and VLT Survey Telescope (VST) imaging for our cluster sample.

From our classification scheme in Section \ref{sec:spectral_results}, we find that Seyfert objects make up 17$\%$ of the misaligned sample in this study. Given the energetic feedback from the AGN at the centre of such objects, we consider the possibility that their observed misalignments are due to outflowing gas. Furthermore, the optical imaging available for our field/group misaligned galaxies allows us to identify clear signs of mergers and tidal interactions.
We first want to separate misalignments potentially caused by outflows from those caused by new gas being accreted.
For cases where gas has an external origin, we try to separate between mergers, tidal interactions (which are
not clear mergers) and systems that do not show any clear signs of gravitational interaction. We will refer to
this last category as \textit{gas accretion from other sources} (i.e. from a galaxy's outer halo or filaments), noting that objects in this category may still include mergers/tidal interactions that happened in the past and for which now there is no sign left in the optical data at our disposal.

We identify galaxies with clear post-merger/merger remnant signatures in our misaligned sample using the criteria outlined in \cite{Barreraballesteros2015}.
Tidally interacting/close-pair objects are identified as having asymmetries in both the optical and kinematic maps (e.g. tidal tails, bridges, warps, plumes), as well as a companion object displaying similar features (e.g. the third row of Figure \ref{fig:Galaxy_maps_examples}). While the mergers and tidally interacting galaxies might have similar morphological tracers, a galaxy is classified as the latter only if two distinct nuclei are present in the optical image.  
The survival time of merger features was found by \cite{Ji2014Mergers2} to be in the range $\sim2.1-5.8$ Gyr for a brightness limit of 25 mag $\rm{arcsec^{-2}}$ and $\sim4.0-8.3$ Gyr at 28 mag $\rm{arcsec^{-2}}$, for mass ratios ranging from 1:10 to 1:1, in both isolated and cluster environments. While the surface brightness limit of the HSC imaging used in this work is $\sim 27.8\ \rm{mag\ arcsec^{-2}}$  in the \textit{g} band (\citealt{HSC_LSBG}), this threshold for the VST and SDSS imaging is only $\sim$25 - 25.5 $\rm{mag\ arcsec^{-2}}$ (see e.g. \citealt{Arnaboldi1999,Ji2014Mergers2}). As a result, for the misaligned cluster galaxies where SDSS or VST imaging is used, our study is expected to identify past mergers (with mass ratios above 1:10, the minimum value probed by \citealt{Ji2014Mergers2}) and tidal interaction events up to $\sim2.1-5.8$ Gyr old. If mergers with mass ratios below 1:10 were able to produce global stellar-gas kinematic misalignments, we cannot exclude the possibility of such features not being probed in misaligned galaxies where SDSS or VST imaging is used. Notably however, only 18/169 misaligned objects are in this category. 

After the identification of objects with merger and tidal interaction signatures in our sample, we employ the following method to classify misaligned galaxies according to the most-probable physical cause of the kinematic decoupling:

\begin{itemize}
    
    \item \textbf{Outflows:} galaxies spectrally classified as Seyfert objects, with kinematic offset angles in the range $[30^{\rm{o}}-150^{\rm{o}}]$ (i.e. non counter-rotating) are classified as having misalignments due to outflowing gas;
    
    \item \textbf{Mergers and tidal interactions/close pairs:} galaxies with clear merger or tidal interaction signatures, which have not been identified as having outflow-driven misalignments, are classified as having the respective process (merger or tidal interaction) as the cause of the kinematic decoupling;
    
    \item \textbf{Gas accretion from other sources:} Galaxies which do not fall within the two categories above are classified as having 
    misalignments due to gas accretion from other sources.
    
\end{itemize}

Although AGN-driven outflows are believed to be preferentially oriented perpendicular to the galactic disc, following the path of least resistance (e.g. \citealt*{TilmanHartwig2018}), observational studies have reported outflows more closely aligned with the stellar component (e.g. \citealt{Barbosa2009}; \citealt{Husemann2019}), which have also been recovered in simulations (e.g. \citealt{Tanner2022}). While counter-rotating stellar-gas configurations (in projection) have been previously interpreted as outflow-driven (\citealt{RongxinLuo2019L}), such scenarios have not been recovered in simulations. We therefore make the conservative choice of only considering outflows to produce observed stellar-gas misalignment angles in the range  $[30^{\rm{o}}-150^{\rm{o}}]$. The median $|\Delta PA_{\rm{stars-gas}}|$  of our outflow-driven misaligned galaxies is $63^{\rm{o}}$ ($102^{\rm{o}}$ for all the Seyfert objects). We visually inspect all such galaxies in our sample, noting than one of these objects (CATID 618993) displays clear signs of having suffered a head-on collision which shock-heated the gas in the central parts, resulting in the observed spectral properties. The galaxy is tidally interacting with a nearby companion and is therefore re-classified as such.

Our outflow classification is based on the assumption that, given the energetic feedback from the AGN at the centre of Seyfert galaxies, misalignments in such objects are unlikely to have been caused by smooth gas accretion. This scenario has been recovered in cosmological hydrodynamic simulations, e.g. \cite{Nelson2015} who found that when feedback is considered, gas accretion from the intergalactic medium in the `smooth mode` is significantly suppressed. Furthermore, \cite{Vandevoort2011} found that AGN feedback prevents gas that enters haloes in the hot mode from collapsing onto the central galaxy. Finally, while \citet{Ho2016b} found evidence of star formation-driven galactic winds generating outflows in SAMI galaxies, we note that all such galaxies have been found to have globally aligned stellar-gas rotation. 

As expected, none of the galaxies classified as tidally interacting in our sample have counter-rotating configurations (their stellar-gas kinematic offset angles are $<92^{\rm{o}}$ in all cases). We note that in this work, we place the tidally interacting galaxies into the category of misalignments caused by gas accretion (i.e. gas exchange between the interacting galaxies). While we cannot exclude the possibility that the gravitational interaction alone (or indeed another process, e.g. gas accretion from another source) is the cause of the misalignment, we note that one of the five such galaxies in our sample (CATID 22633) displays a clear evidence of gas exchange with a close-by, lower-mass, more tidally disturbed companion in the optical image. Notably, this assumption does not affect the results described in the next section, given the small number (5) of tidally interacting galaxies in our sample. 

\cite{Bryant2019} considered the possibility of kinematic misalignments in the cluster environment being  due to gas stripped from the disc via hydrodynamic effects, as a galaxy falls through the intra-cluster medium. Such a process will produce a gradient in the gas velocity field in the direction of infall. However, after careful inspection of our data, we find only one galaxy (<1\%) supporting a ram-pressure stripping origin for the misalignment. This galaxy (CATID 9008500100) has been previously identified by \cite{Owers2019} as having one-sided/extra-planar H$\alpha$ emission. None of the other SAMI cluster galaxies with such properties in \cite{Owers2019} have been found to be globally misaligned here.

\begin{figure}
	\centering
	\includegraphics[width=\columnwidth]{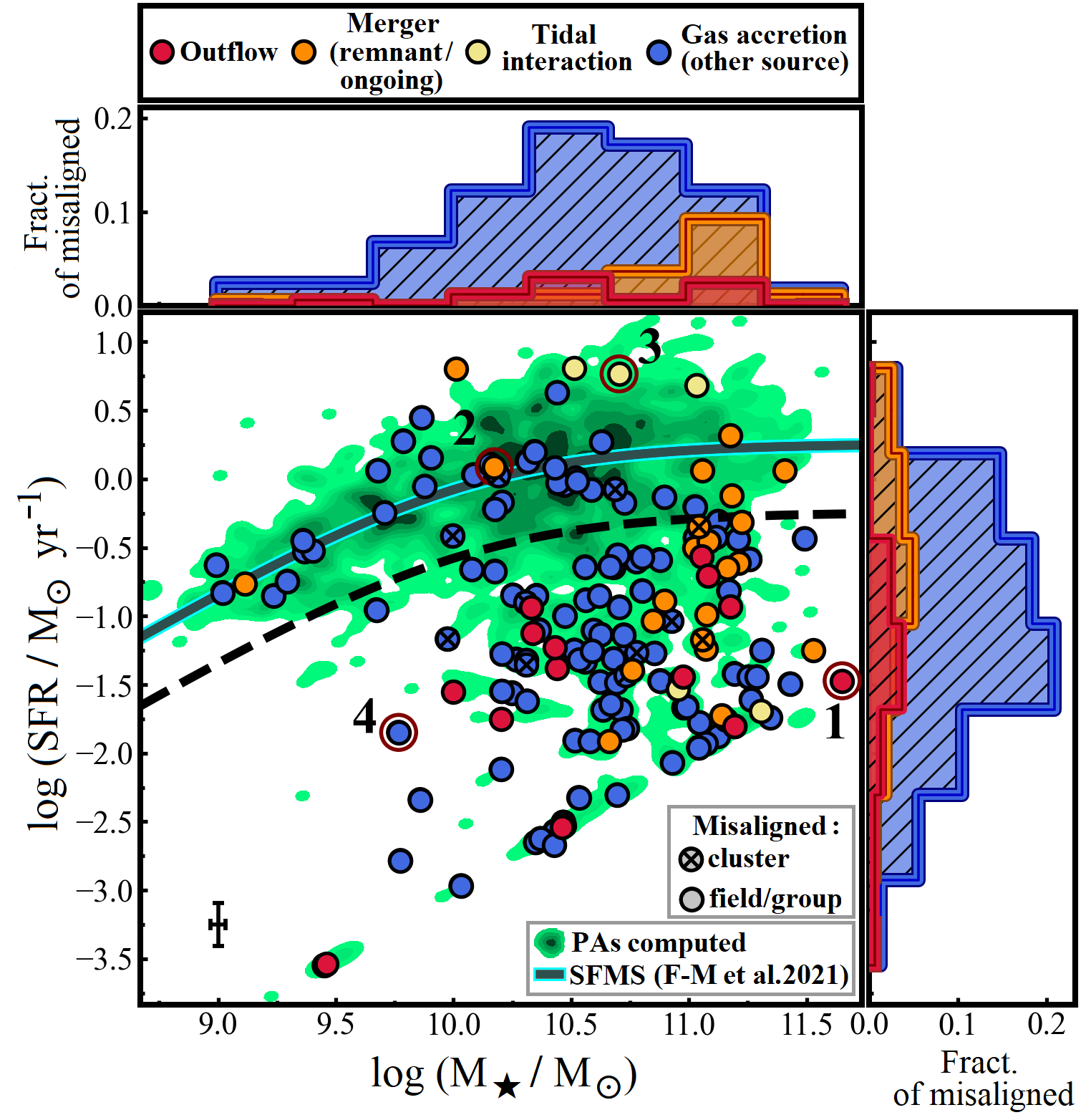} 
    \caption{SFR vs stellar mass for the misaligned (colored circles) and parent samples (green contours). The colour of each misaligned galaxy reflects the most probable physical cause of the misalignment (see Section \ref{sec:misalign_cause}). Circles with an 'x' marker depict cluster galaxies in our misaligned sample. The teal-cyan solid line and error bars are the same as in Figure \ref{fig:Mstar_SFR_nsersic_plots}. The circled galaxies are shown in Figure \ref{fig:Galaxy_maps_examples}, as indicated by the numbers (1-4). The adjacent histograms show the distribution of misaligned galaxies in SFR and $M_{\star}$, separated by the most probable cause of the misalignment (with tidal interactions and mergers combined in the orange histogram). The one cluster galaxy classified as having gas stripping causing the misalignment does not have a SFR measurement and is not shown here.}
    \label{fig:Misalignment_cause_plot}
\end{figure}

The results of our classification are presented in Figure \ref{fig:Misalignment_cause_plot}, with misaligned galaxies shown on the $M_{\star}$ - SFR plane, color-coded according to the main four physical causes of misalignments identified in this work, as described above: outflows (8\%), tidal interactions (3\%), recent or ongoing mergers (14\%) or accretion from other sources (74\%), with one other cluster galaxy (<1\%) being identified as having a misalignment due to gas stripping. These results are also presented in Table \ref{tab:tablemaster}.
In summary, while this analysis confirms that the bulk of our misaligned sample is consistent with a gas accretion origin, we find a non-negligible fraction of galaxies for which gas being expelled in an outflow is more likely to be the cause of the kinematic decoupling.

\begin{figure*}
	\centering
	\includegraphics[width=\linewidth]{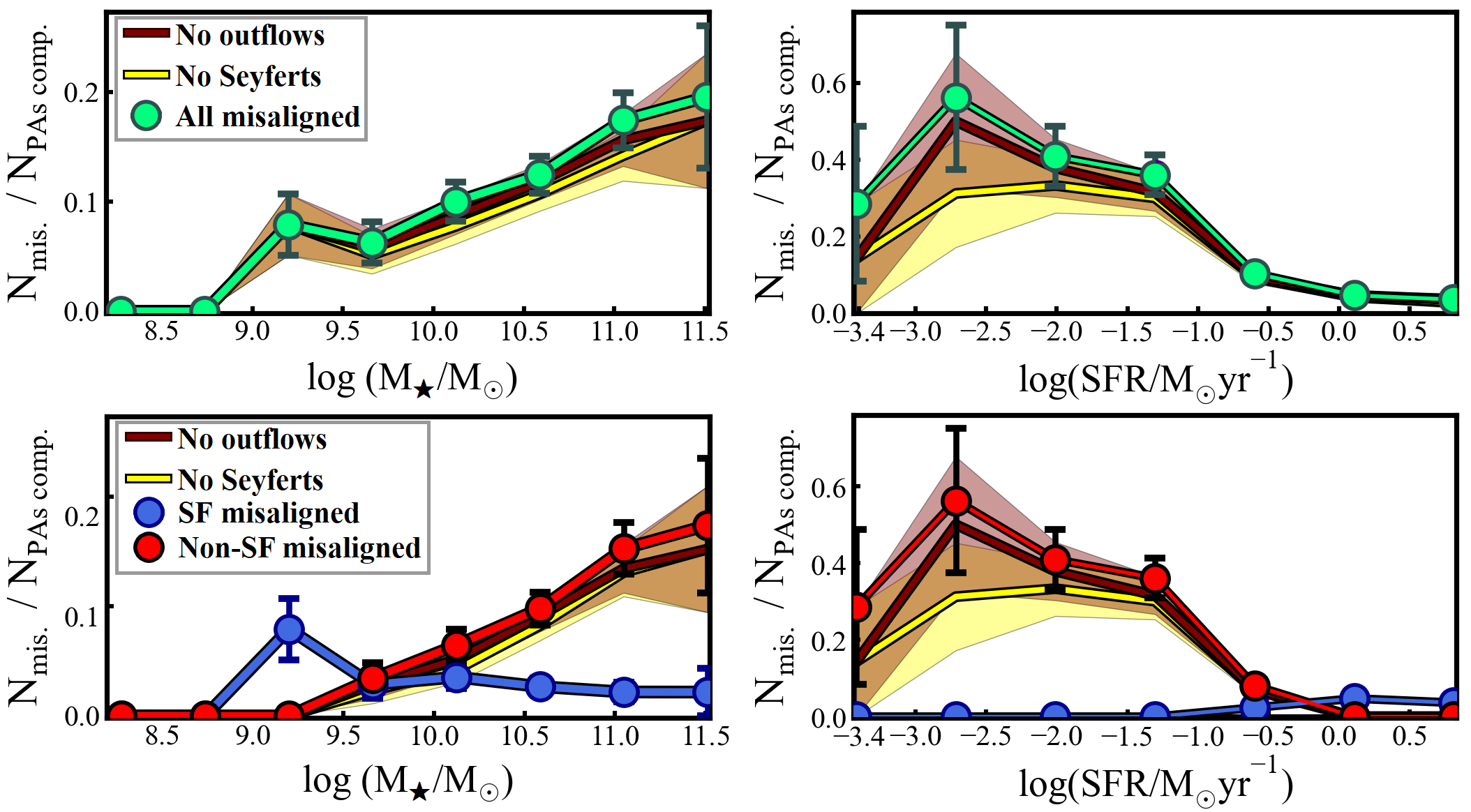}
    \caption{\textbf{(Top row) }Fraction of parent sample galaxies that are misaligned, as a function of stellar mass (\textbf{left}) and SFR (\textbf{right}). The brown and yellow lines show the fractions when excluding the galaxies classified as having misalignments due to outflows, and those spectroscopically classified as Seyfert objects, respectively. The error bars and shaded areas reflect the Poisson uncertainties in the fractions, given the number of galaxies in each bin. \textbf{(Bottom row)} The contribution to the fraction of misaligned galaxies in each bin from SF (blue) and non-SF (red) galaxies, respectively.  }
    \label{fig:Fraction_plots}
\end{figure*}


\setlength{\tabcolsep}{2.2pt} 
\renewcommand{\arraystretch}{1.3}
\begin{table*}
\centering
\caption{Number of misaligned galaxies (including values without outflow-driven cases) and percentage of the sample in each spectral category (SF, LINER, Seyfert - Section \ref{sec:spectral_results}), and split by the most probable physical cause of the misalignment (Section \ref{sec:misalign_cause}).}\label{tab:tablemaster}
\begin{tabular*}{\textwidth}{@{\extracolsep{\fill}}c|ccc|ccccc} 
\hline
\multirow{2}{*}{Number (\%) of}      & \multicolumn{3}{c|}{Spectral class} &  \multicolumn{5}{c}{Misalignment cause}  \\
                              & SF & LINER & Seyfert & Outflow                        & Gas stripping & Merger &  Tidal interaction & Gas accretion                               \\ 
\hline
Misaligned galaxies                  & $37\substack{+7 \\ -3}$ ($22\substack{+4 \\ -2} \%$) & $101\substack{+3 \\ -4}$ ($61 \pm 2 \%$) &  $28 \pm 3$ ($17 \pm 2 \%$) & 14 (8$\%$)   & 1 (<1$\%$)  &  24 (14$\%$) & 5 (3$\%$)   & 125 (74$\%$)                                                       \\
\textit{Non-outflow} misaligned galaxies & $37\substack{+7 \\ -3} (24\substack{+5 \\ -2} \%)$  & $ 101\substack{+3 \\ -4} (67\substack{+2 \\ -3} \%)$     & $14 \pm 3 (9 \pm 2 \%)$                    & - & - & - & - & - \\                                                     
\hline
\end{tabular*}
\end{table*}



\subsection{Misalignment fractions dependency on stellar mass and SFR}
\label{sec:fractions}

The results in Section \ref{sec:physical_properties} suggest that misaligned galaxies generally populate the high-$M_{\star}$, low-SFR region of parameter space.
Figure \ref{fig:Fraction_plots} shows the fraction of parent sample galaxies which are misaligned with respect to stellar mass (left column) and SFR (right column). The bottom row shows the contribution to the observed fractions from SF and non-SF misaligned galaxies, respectively. Our results are consistent with an increase in misalignment fractions with $M_{\star}$ between $\sim 10^{9}\rm{M_{\odot}}-10^{11}\rm{M_{\odot}}$, while our number statistics outside this range do not allow us to discern between different trends (Figure \ref{fig:Fraction_plots} - top left). The observed dependency of misaligned fractions on $M_{\star}$ is driven by non-SF galaxies above $\sim10^{9.5}\rm{M_{\odot}}$, with  the fractions of misaligned SF galaxies being approximately constant across the probed stellar mass range, within uncertainties.

Our findings show increasing misalignment fractions going to lower SFRs, in the range  $0.5 \lesssim$ log(SFR / $\rm{M_{\odot}\ yr^{-1}}$) $\lesssim -2$, while at lower SFR our results become highly uncertain due to low number statistics (Figure \ref{fig:Fraction_plots} - top right). This trend is unsurprisingly driven by the non-SF population below $\sim 10^{-0.5}\rm{M_{\odot}\ yr^{-1}}$, and by the SF sub-sample above this value. 

We also evaluate the contribution of outflow-driven cases and Seyfert objects to observed misalignment fractions. We note that considering only the accretion-driven kinematically decoupled galaxies (i.e. excluding outflows - light red curve in Figure \ref{fig:Fraction_plots}) or the non-Seyfert objects (dark red curve) does not change the observed trends with $M_{\star}$ and SFR, but rather only produces systematically lower fractions. Therefore, the observed trends of misaligned fractions with $M_{\star}$ and SFR are shaped by accretion-driven misalignments and are not significantly affected by AGN-driven outflowing gas.

\subsection{Morphology, stellar kinematics and gas content}
\label{sec:distributions_results}

Previous studies of stellar-gas kinematic misalignments have suggested that the prevalence of such features is correlated with the shape of the stellar mass distribution (e.g. \citealt{Davis2011}; \citealt{CdpLagos2015}; \citealt{Bryant2019}; \citealt{Duckworth}; \citealt{Khim2020I}), the kinematic morphology (i.e. the ratio of ordered to disordered motion, $v/\sigma$; e.g. \citealt{Xu}; \citealt{Khim2020II}) and the cold gas fraction (\citealt{CdpLagos2015}; \citealt{Duckworth}; \citealt{Khim2020I}) of a galaxy.
In this picture, newly accreted misaligned gas takes longer to settle into a stable configuration in galaxies with early-type morphologies or lower pre-existing gas content. 

In this work, we study the interplay between galaxy morphology \& sSFR (used here as our best proxy for cold gas content), and the phenomena of gas accretion and dynamical settling into stable configurations. Since both properties are interconnected, we attempt to determine whether they are both playing a major role in driving the timescales of misalignments, or if there is a dominant mechanism. To this end, galaxies where outflows or gas stripping have been identified as likely to be causing the kinematic decoupling (15 out of 169 objects) are excluded from this analysis.

\begin{figure*}
	\centering	\includegraphics[width=\linewidth]{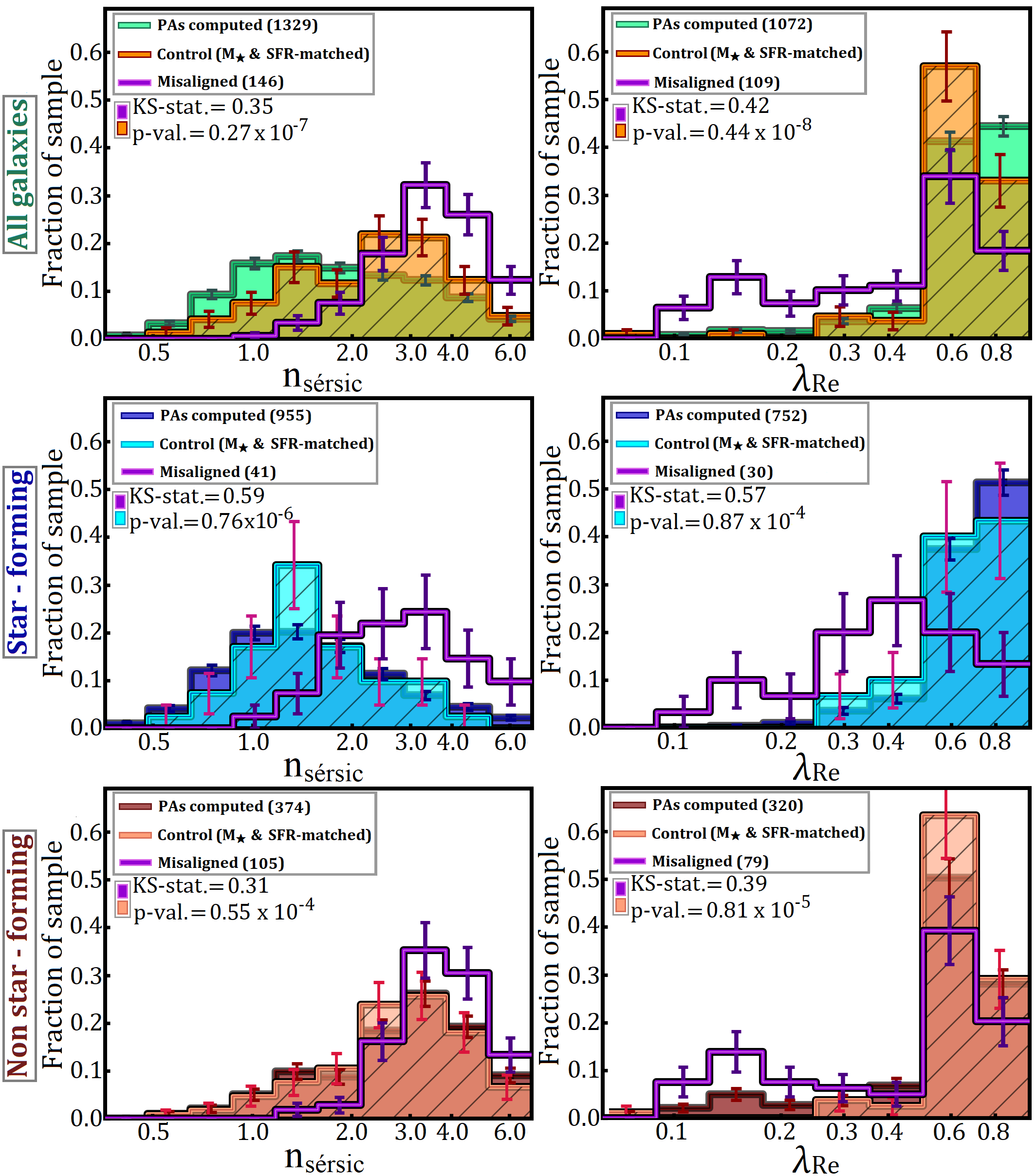}    \caption{Distributions of the log of S\'ersic indices ($n_{\rm{S\Acute{e}rsic}}$ - \textbf{left column}) and spin parameters ($\lambda_{Re}$ - \textbf{right column}), with the parameter axes values showing the corresponding (non-logged) $n_{\rm{S\Acute{e}rsic}}$ and $\lambda_{Re}$ values. The \textbf{top} row displays the distributions for the whole parent (green) and misaligned (purple) samples. The orange histograms show control samples of aligned galaxies matched in $M_{\star}$ and SFR to the misaligned galaxies (and of the same size), drawn from the whole parent sample. The \textbf{middle} and \textbf{bottom} rows show the distributions for the SF (blue) and non-SF (red) sub-samples of the parent (dark blue/red) and misaligned (purple) populations. The light blue and red histograms show control samples of aligned SF and non-SF galaxies matched in $M_{\star}$ and SFR to the misaligned galaxies (and of the same size), drawn (only) from the respective SF/non-SF parent sub-samples. 
	The number of galaxies in each population is displayed in the legend. The KS-statistics and p-values correspond to a statistical comparison of the control and misaligned samples in each panel. The error bars reflect the Poisson uncertainties given the number of galaxies in each bin.  }
    \label{fig:nsersic_lambda_re_dist}
\end{figure*}

First, we study the interplay between gas accretion and galaxy morphology/stellar kinematics by analysing the distribution of accretion-driven misaligned galaxies with S\'ersic index ($n_{\rm{S\Acute{e}rsic}}$) and stellar spin parameter ($\lambda_{Re}$), in comparison to those of the aligned parent sample galaxies, at fixed stellar mass and SFR. 
To match our misaligned sample in $M_{\star}$ and SFR, for each kinematically decoupled object we find the aligned parent sample galaxy  with the closest $M_{\star}$ and SFR values, i.e. that minimises the sum of the differences in the two parameters ($|\Delta \rm{log}(M_{\star})_{\rm{misaligned-aligned}}|+|\Delta \rm{log}(SFR)_{\rm{misaligned-aligned}}|$). An aligned parent sample galaxy can only be matched with one single misaligned object. 
These distributions are presented in Figure \ref{fig:nsersic_lambda_re_dist}, where the top row shows the whole misaligned (purple) and parent (green) samples, together with the relevant control sample (orange), while the middle and bottom rows display the same results when the match is done for SF and non-SF galaxies, separately. As our misaligned sample is dominated by passive systems, matching SF and non-SF populations individually allows us to establish whether the same conclusions can be reached for both samples, independently. The left and right columns show the distributions in $n_{\rm{S\Acute{e}rsic}}$ and $\lambda_{Re}$ for galaxies where measurements of each parameter are available. 

Controlling for $M_{\star}$ and SFR has a larger effect when considering the whole misaligned and parent distributions, as is reflected in the difference between the green (parent sample) and orange (control sample) histograms in the top row of Figure \ref{fig:nsersic_lambda_re_dist}. When splitting our samples into SF and non-SF, the parent and misaligned sub-samples are expected to be more closely matched in SFR as they occupy the same parameter space around the SFMS and thus the matching has a smaller effect. The \textit{misaligned} and \textit{control} sample distributions with $n_{\rm{S\Acute{e}rsic}}$ and $\lambda_{Re}$ were compared using a KS-test, as displayed on each panel of Figure \ref{fig:nsersic_lambda_re_dist}. Our results show that distributions of misaligned and (SFR- \& $M_{\star}$-matched) aligned galaxies are statistically different in all cases. Kinematically decoupled galaxies have larger S\'ersic indices and lower stellar spin than aligned objects with similar distributions in stellar mass and SFR. The differences are reduced when considering the SF and non-SF populations individually (middle and bottom rows of Figure \ref{fig:nsersic_lambda_re_dist}). We note, however, that the number statistics are reduced when looking at trends with the stellar spin parameter since misaligned galaxy distributions with $\lambda_{Re}$ typically contain $\sim$ 73-75 \% of the total sample, while for the parent sample, the fraction is in the range $\sim$ 79-86 \%. Overall, this result shows that morphology (visual and kinematic) is a key property associated with misalignment prevalence both above and below the SFMS.

Second, we look at the effect of sSFR by performing a similar analysis and controlling for stellar mass and morphology ($n_{\rm{S\Acute{e}rsic}}$). The results are shown in Figure \ref{fig:sSFR_Dist}, where we split our misaligned and parent samples into late- and early-type galaxies (middle and bottom respectively - see section \ref{sec:physical_properties}).
We note that an equivalent matching in $\lambda_{Re}$ is not possible here, given that reliable measurements for the stellar spin parameter are not available for a significant portion of the misaligned sample.
The only case when the misaligned and control sample distributions are statistically similar is for the late-type sub-sample (Figure \ref{fig:sSFR_Dist} - bottom), with the caveat that only 25 of our (non-outflow) kinematically decoupled galaxies are in this morphological classification. If confirmed to be significant, this would suggest that sSFR is correlated with misalignment occurrence only for the early-type galaxy population.

As shown in Appendix \ref{A1}, our conclusions are unaffected if we remove counter-rotating galaxies (which are likely already stable systems, within errors) from the analysis.  In summary, our results indicate that the likelihood of showing misalignments depend on both SFR and morphology. As discussed in Section \ref{sec:dpa_distr_discuss}, this is a critical point for understanding the origin of our misaligned sample.

\section{Discussion}
\label{sec:discussion}

In this section, we begin by discussing the implications of the position angle offset ($|\Delta PA_{\rm{stars-gas}}|$) distributions of different galaxy populations for the timescale of gas settling, as well as the influence of morphology and sSFR on driving misalignment timescales (Section \ref{sec:dpa_distr_discuss}). 
Following this, we discuss our spectral classification results for kinematically decoupled galaxies in the context of gas accretion, as well as the contributions of different physical processes to misalignment numbers. Finally, the effect of stellar mass and SFR on driving the fractions of kinematically decoupled galaxies is analysed. These aspects are presented in Section \ref{sec:discuss_spectral_phys_cause}. 

\subsection{Galaxy properties associated with misalignments - the effect of morphology and star formation}
\label{sec:dpa_distr_discuss}

\subsubsection{Misalignment fractions}
The total fraction of misaligned galaxies ($11.7\substack{+2.7 \\ -0.9}$\%) agrees well with the estimates of \cite{Bryant2019} ($11\pm 1$\%) for the previous data release of SAMI. Past studies based on the MaNGA Survey found slightly lower misalignment fractions, for example \cite{Xu} ($\sim$6.6 \%) and \cite{Duckworth} ($\sim$9.2 \%), across all morphological types, with the mention that the former study excludes all non-rotators, on-going mergers or broad-line AGN from their misaligned sample. Excluding all such objects from our sample (as well as tidally interacting galaxies) reduces our fraction of kinematically decoupled objects to $\sim$7.7 \%.

When splitting by morphology using a cut at $n_{\rm{S\Acute{e}rsic}}$ = 2.5, we found that $30 \substack{+3 \\ -2}\ $\% of galaxies with early-type morphologies are misaligned. This finding is again consistent with the fraction determined by \cite{Bryant2019} across all environments ($\sim$33 $\pm$ 5 \%) as well as the results of \cite{Davis2011} based on the $\rm{ATLAS^{3D}}$ project, who found that a fraction of $\sim 36\pm 5$ \% fast-rotating early-type galaxies (in both the field and cluster environments) have misaligned stellar and ionised gas rotation.

\subsubsection{Comparison of PA offset
distributions and the effect of morphology $\&$ sSFR on driving misalignment timescales}
\label{sec:discuss_misalignment_dPA_timescales}
The $|\Delta PA_{\rm{stars-gas}}|$ distribution encodes information about the relaxation, depletion and accretion timescales of gas. In our current understanding of the evolution of accretion-driven misaligned configurations, the newly introduced gas has an angular momentum vector that leads it to form a rotating disc, setting the initial PA. The gravitational potential of the stellar component exerts a torque on this disc, causing it to settle into the same plane, into a co- or counter-rotating configuration \citep*{Tohline1982}. Furthermore, any pre-existing gas reservoir (i.e. present in the galaxy before the accretion event) will exert a viscous drag on the newly acquired gas, speeding up the alignment process \citep{Davis2011}. Our results in Section \ref{sec:dpa_distribution} seem to confirm that both the shape of the stellar mass distribution and the pre-existing gas content play a dominant role in regulating the misalignment process. We tested the effect of excluding all outflow-driven misalignments and Seyfert objects from our sample (Appendix \ref{A2} - Figure \ref{fig:A2}), noting that this did not change the statistical significance of our results regarding $|\Delta PA_{\rm{stars-gas}}|$ distributions in Section \ref{sec:dpa_distribution}.

The fraction of parent sample galaxies displaying kinematically decoupled configurations still in the process of stabilising (i.e. not counter-rotating) is significantly higher for the non-SF and early-type sub-samples (20\% and 19\%) than for the SF and late-types (1.7\% and 1.0\%). There are $ 12 \substack{+8 \\ -7}$ times more misaligned galaxies in an unstable configuration in the non-SF population than in the SF, and $ 19 \substack{+9 \\ -13}$ times more in the early-type sub-sample than in the late-type. The errors in these fractions are significant, especially at the lower end, given the small number statistics in the SF/late-type  misaligned populations. Within these conservative limits, the difference between the late- and early-type populations is consistent with the value found by \cite{Bryant2019} ($15\pm 7$), with the mention that our morphological classifications is based on S\'ersic index, while \cite{Bryant2019} separate their populations based on visual morphology. These results suggest that misaligned configurations have significantly longer timescales in galaxies with a more centrally concentrated stellar mass distribution, or alternatively in objects with low SFR (i.e. more gas poor).

If we consider the effect of morphology/stellar distribution alone, we can follow \cite{LakeNorman1983} who proposed that the relaxation time of a misaligned gas disc due to the torque of the stellar component is a fraction $1/\epsilon$ (where $\epsilon$ represents the ellipticity) of the dynamical time (i.e. the time taken for a full precession of the gas disc). This estimate agrees relatively well with the result of \cite{Freekevdv2015}, who found that the gas settling time in a misaligned early-type galaxy is $\sim 6$ dynamical times (after accretion rate has dropped significantly).

Similarly, \cite{Bryant2019}, following on from \citet{Tohline1982}, suggest that the dynamical torque exerted by the stellar component depends on ellipticity, gas angular velocity and initial misalignment angles. 
Considering the typical range of ellipticities for our sample (0.16-0.74), this implies that the effect of morphology alone would result in fractions of unstable misalignments up to $\sim$3.2 times higher in early-types compared to late-types (considering the extremes of the ellipticity range). However, in our sample, we find that the difference in misalignment fraction between early-types and late-types is several factors higher, albeit with a significant uncertainty ($ 19 \substack{+9 \\ -13}$). These results suggest that morphology is not the only galaxy property driving the timescale of gas settling.
Thus, part of the surplus of misaligned galaxies in unstable configurations for the non-SF/early-type populations compared to the SF/late-type sub-samples can be explained by considering the viscous drag forces between accreted and pre-existing gas, which speeds up the process of alignment. In this picture, galaxies which have larger gas reservoirs before a misaligned accretion event will tend to stabilise over shorter periods of time. 

The need for both morphology and gas content/star formation playing a simultaneous role in setting the fraction of misalignments is reinforced by the analysis presented in Section \ref{sec:distributions_results}. There, we have shown that misaligned systems have higher $n_{\rm{S\Acute{e}rsic}}$ and lower $\lambda_{Re}$ than aligned galaxies at fixed $M_{\star}$ and SFR, as well as lower sSFR at fixed $M_{\star}$ and $n_{\rm{S\Acute{e}rsic}}$. When separating into different sub-populations, our results show that morphology is significantly associated with misalignment prevalence both above and below the SFMS, while sSFR might only have an important role for the early-type population.
Interestingly, the same result is recovered if we focus only on misaligned galaxies in unstable configurations (non counter-rotating), as shown in Appendix \ref{A1} (Figure \ref{fig:A1}), suggesting that these galaxy properties are not only associated with the likelihood of misalignment, but also have a dominant effect on the timescale for gas precession in a stable configuration.

Our results are qualitatively consistent with those found in previous studies of stellar-gas kinematic misalignments which addressed the correlation between morphology/stellar mass distribution (\citealt{Davis2011}; \citealt{CdpLagos2015}; \citealt{Bryant2019}; \citealt{Duckworth}; \citealt{Khim2020I}; \citealt{Zhou2022}), gas fraction (\citealt{CdpLagos2015}; \citealt{Duckworth}; \citealt{Khim2020I}), and the likelihood and timescales of kinematically decoupled features. The physical implications of our findings can be formulated considering our arguments regarding the timescale for the settling of a misaligned gas disc in a stable configuration: the gravitational torque exerted by the stellar disc on the misaligned gas component is lower for more centrally concentrated stellar mass distributions (i.e lower ellipticity), resulting in kinematically decoupled configurations having larger timescales in galaxies with early-type morphologies. Furthermore, in a misaligned accretion event, the newly acquired gas interacts dynamically with the gas reservoir already in the galaxy, with viscous drag forces between the two speeding up the settling process of accreted gas in a stable configuration. Such a scenario has been recovered in the toy model presented by \cite{Khim2020II} for the relaxation of gravitationally self-interacting rings of stars and gas, which considers both gravity and viscous drags between different gas components. 

Alternatively, preferentially aligned accretion in late-type galaxies \citep{Bryant2019} would lower the average gravitational torquing timescale. It has been shown by previous work based on the Galaxies-Intergalactic Medium Interaction Calculation (GIMIC) \citep{Sales2012} and Horizon-AGN \citep{Welker2017} simulations that disc galaxies tend to form predominantly from aligned gas accretion, whereas more spheroidal morphologies are the result of accretion episodes in which the angular momentum of the infalling gas is misaligned with respect to that of the host galaxy. In this picture, the fact that we observe significantly fewer misaligned late-type galaxies than early-types would simply be the result of the fact that the morphology of late-types is the product of predominantly aligned accretion. 

\begin{figure}
	\centering \includegraphics[height=18.9cm]{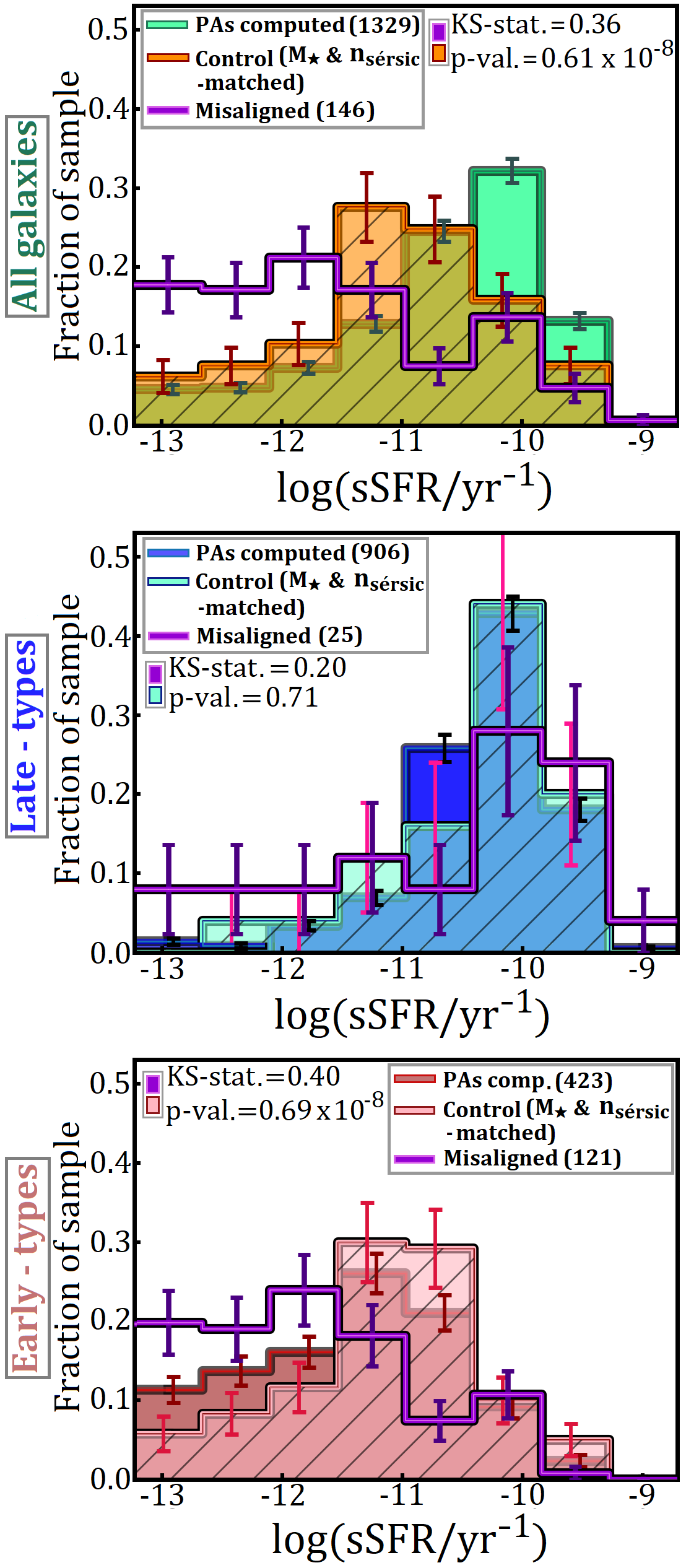}   \caption{\textbf{(Top)} Distributions of log(sSFR) for the whole misaligned (purple) and parent (green) samples, as well as a control sample of aligned galaxies (orange) matched in $n_{\rm{S\Acute{e}rsic}}$ and $M_{\star}$ to the misaligned one, and of the same size. \textbf{(Middle \& Bottom)}. The same as the top plot, for the late- and early-type  parent (dark blue/red) and misaligned (purple) sub-samples, respectively. The light blue and red histograms show control samples matched in $n_{\rm{S\Acute{e}rsic}}$ and $M_{\star}$ to the misaligned galaxies (and of the same size). The galaxies in these control samples are drawn only from the respective late/early-type parent sub-samples. The number of galaxies in each sample is displayed in the legend. The KS-statistics and p-values correspond to a statistical comparison of the control and misaligned samples in each panel. The error bars reflect the Poisson uncertainty given the number of galaxies in the respective bin.}
    \label{fig:sSFR_Dist}
\end{figure}

While our results for the entire misaligned sample show that both $n_{\rm{S\Acute{e}rsic}}$ and sSFR have significant contributions in driving misalignment timescales, the marginal evidence (given the low number statistics) of sSFR being less significant than morphology for the late-type misaligned population (Figure \ref{fig:sSFR_Dist} - middle) is consistent with the framework in which disc morphologies are largely the result of aligned gas accretion (\citealt{Sales2012}; \citealt{Welker2017}), which would act to explain the low numbers of misalignments in this population. If such results were confirmed with a more statistically powerful sample of misaligned late-type galaxies (including passive systems), it would suggest that the stellar-gas misalignment phenomenon is driven by an interplay between three key elements: \textbf{(1)} morphology (shape of the stellar mass distribution); \textbf{(2)} pre-existing gas content influencing the settling timescale of accreted, kinematically decoupled gas in a stable configuration; \textbf{(3)} late-type morphologies in the nearby Universe being the result of preferentially aligned gas accretion at higher redshifts. We note, however, that our study only provides significant observational evidence for the contribution of the first two. The extent to which the latter phenomenon contributes to our results can be better understood with a more statistically powerful sample of misaligned late-type galaxies.

\subsubsection{Counter-rotating misalignments}
Our results confirm the statistical significance of the counter-rotating ($150^{\rm{o}}-180^{\rm{o}}$] peak in the $|\Delta PA_{\rm{stars-gas}}|$ distribution, as seen in Figure \ref{fig:dPA_plots}, suggesting that this configuration is stable. This finding is in agreement with the simulation-based work of \cite{OsmanBekki2017}, which found counter-rotating gas discs to be \textit{permanently} stable, i.e. until the gas is consumed by star formation. The existence of this peak has been previously reported by \cite{Bryant2019}, with \cite{Duckworth} only finding a small increase in the number of misaligned galaxies around $180^{\rm{o}}$, which was not present in their matched Illustris TNG100 sample. Works based on the Evolution and Assembly of GaLaxies and their Environments (EAGLE) simulation, looking at the offset between angular momenta of stars and star-forming gas \citep{CdpLagos2015, Casanueva2022} have also failed to find a counter-rotating peak. Furthermore, \cite{Davis2011} found no counter-rotating peak in the kinematic offset angle distribution for a sample of early-type fast-rotating galaxies in the field/group and Virgo cluster, albeit with relatively low number statistics (111 galaxies). This result was interpreted by \cite{Davis2015} as a potential evidence for the fact that the gas depletion timescales from star-formation \citep{Bigiel2011,Kennicutt1998} or destruction timescales from AGN feedback \citep{Hopkins2006} are shorter than previously believed, or alternatively that the settling of misaligned gas occurs over longer periods of time than assumed (e.g \citealt{LakeNorman1983}). Such assumptions are inconsistent with the existence of a significant counter-rotating peak found in our study.

Furthermore, \cite*{Stevens2016} showed that precession of the gas disc angular momentum around that of the stars is required in order to obtain a $|\Delta PA_{\rm{stars-gas}}|$ distribution with a peak in both the aligned and counter-rotating regions. The counter-rotating peak in the \textit{precession-on} mode was found to be $\approx$10 \% of the co-rotating one, while the \textit{precession-off} prescription produced a gradually declining distribution with increasing $|\Delta PA_{\rm{stars-gas}}|$. Our results confirm the importance of gas disc precession due to the gravitational torque of the stellar disc. The $\sim$10:1 size ratio of the co- and counter-rotating peaks found by \cite{Stevens2016} is consistent with our results for the non-SF and early-type populations individually ($\sim$12\%; see also Table \ref{tab:TableA1}); however, for our entire sample of galaxies, the counter-rotating peak is only 4.2\% of the aligned one, while this percentage drops to 2.0\% and 1.4\% for the SF and late-type sub-samples. 

While both the co- and counter-rotating peaks include galaxies with externally accreted gas which has settled due to stellar disc torques, the co-rotating peak also contains cases where the interaction with the pre-existing gas reservoir has exerted a drag on the newly accreted gas, causing it to align. This is the case for accretion events where the angular momentum of the infalling gas is smaller than that of the pre-existing component and acts to align the newly acquired gas into a co-rotating configuration regardless of its initial misalignment angle, i.e. acts on counter-rotating accreted gas as well \citep{Davis2015}. Such a scenario is expected to be more prevalent in SF galaxies with a significant cold gas component, or alternatively in late-type galaxies where any potential previous gas reservoir would have higher angular momentum, in accord with our findings on stable configuration peak size ratios. This effect is not considered in the \cite{Stevens2016} semi-analytic model and could explain the discrepancies of the co- to counter-rotating peak sizes.   

In summary, our findings highlight the important role of the precession of misaligned gas discs and friction between different gas components in shaping $|\Delta PA_{\rm{stars-gas}}|$ distributions, and appear to rule out unusually short gas depletion or destruction timescales previously reported (\citealt{Davis2015}).

\subsection {Properties of misaligned galaxies and the contribution of different physical processes to their creation}
\label{sec:discuss_spectral_phys_cause}
 
\subsubsection{Ionisation mechanisms in misaligned galaxies}

We performed a spectral classification of misaligned galaxies in our sample (Section \ref{sec:spectral_results})  based on the [\ion{N}{II}]- and [\ion{S}{II}]-BPTs (Figure \ref{BPT_plots}) in order to identify the main gas ionisation mechanisms in these objects. Our results show that only $22\substack{+4 \\ -2}$ \% of kinematically decoupled galaxies have ionisation dominated by star formation, while the majority ($61 \pm 2$ \%) are classified as LINERS. From the remaining 28 galaxies with a significant AGN component (Seyfert objects), 14 were classified as having their misalignment potentially driven by an outflow (Section \ref{sec:misalign_cause}), leaving a total of 155 galaxies where gas accretion (from any source) is the most probable cause of the misalignment feature. Out of these, only $37\substack{+7 \\ -3}$ objects, corresponding to $24\substack{+4 \\ -2}$ \%, have their ionisation dominated by star-formation ($26\substack{+5 \\ -2}$ \% of all non-Seyfert misaligned galaxies galaxies). This result suggests that the newly accreted gas is not fuelling new star formation in most ($76\substack{+2 \\ -4}$ \%) cases, a picture which is qualitatively consistent with the distribution of SFR for misaligned galaxies in Figure \ref{fig:Mstar_SFR_nsersic_plots} (left). While these numbers refer to the dominant ionisation mechanism within 1 $R_{\rm{e}}$, as discussed in Section \ref{sec:spectral_results}, our findings should be statistically representative for the entire misaligned gas component.

\subsubsection{Census of physical processes causing misalignments}

In this work, we combined stellar-gas position angle offset measurements with spectral properties and optical imaging of misaligned galaxies to identify the most probable physical cause of the kinematically decoupled configuration. This classification was presented in Section \ref{sec:misalign_cause} and shows that stellar-gas misalignments in our sample are caused by: galactic outflows (8 \%); recent or ongoing gas-rich mergers (14 \%); tidal interactions (3 \%); gas accretion from other sources such as outer halo, filaments, or a past merger (74 \%), and ram-pressure stripping (one galaxy, < 1 \%).

The result of our classification of physical processes causing misalignments is shown in Figure \ref{fig:Misalignment_cause_plot} for the four main formation channels identified. We note that cases of on-going mergers, merger remnants and tidal interactions (orange and yellow points) are typically found at high stellar masses ($\sim 90$ \% of cases above $10^{10.5}\ \rm{M_{\odot}}$), suggesting that such objects are predominantly tracing accretion from a major gas-rich merger and that more minor merging events are less likely to produce a global PA offset. We note, however, that this result is expected to be influenced to a small extent by the fact that merger and interaction signatures are less visible in the optical images of low-$M_{\star}$ galaxies. Outflow-driven misaligned galaxies have SFR values lower than -0.78 dex below the SFMS for their corresponding $M_{\star}$ in all cases, as expected given the energetic AGN feedback in these objects. 

While our assumption that counter-rotating configurations cannot be the result of outflowing gas is physically warranted, given our measurement errors for kinematic PAs, the total number of Seyfert objects in our misaligned sample (17 \%) provides an upper limit on the contribution of outflows to causing kinematically decoupled configurations. Furthermore, the identification of merger- and interaction-driven misalignments in this work gives an absolute lower limit on the contribution of gas accretion from these processes towards causing a stellar-gas kinematic decoupling in the nearby Universe. We cannot exclude the possibility that a number of misalignments which have not been identified as either mergers, tidal interactions or outflows have been caused by accretion from a past merger which is no longer visible in the optical image. The time taken for gas settling in a simulated elliptical galaxy initially misaligned by $120^{\rm{o}}$ was found by \cite{Freekevdv2015} to be $\sim$ 3 Gyrs (including the period of continuous gas accretion). However, this estimate is expected to vary significantly depending on morphology, initial gas content and initial misalignment angle. 
On the other hand, merger features were found by \cite{Ji2014Mergers2} to persist for $\sim$ 2.1 - 7.5 Gyrs, depending on the galaxy's brightness.

We note that large-scale environment appears to play a role in driving stellar-gas misalignments, with only 18/304 (6\%) cluster galaxies in our parent sample being kinematically decoupled, while this is the case for 151/1141 (13\%) galaxies in the field/group. 
Using the group membership classification based on  the GAMA galaxy group catalogue of \cite{Robotham2011} (see also \citealt{vds2021}), we find that 51 (30\%) misaligned galaxies are satellites in groups with halo masses above $10^{13}\ \rm{M_{\odot}}$, while 84 (50\%) are centrals and 34 (20\%) are classified as being isolated objects. Of the 51 misaligned satellite galaxies, 44 (26\% of the total misaligned sample) are classified as being caused by gas accretion from other sources (i.e. not mergers or tidal interactions), which provides a potential upper limit to the number of galaxies in our sample where the misalignment cause could involves environmental processes such as gas stripping in the group/cluster, as this is expected to predominantly affect satellites \citep*{Cortese2021}.

Distinguishing between the possible physical causes of misalignments on a case-by-case basis has only been previously attempted in simulation-based studies, most notably the work of \cite{Khim2020II} who studied a sample of $\sim$27900 galaxies in the Horizon-AGN cosmological simulation, identifying four main channels for misalignment formation: mergers (35 \%), interactions with other galaxies (23 \%), environmental processes (for which the main candidate identified was ram-pressure stripping) \& interactions with central galaxies (21 \%), and secular evolution involving smooth accretion via neighboring filaments (21 \%), with typical uncertainties on these values quoted at 10 \%. While these processes do not entirely coincide with the physical causes identified in this work, it was noted by \cite{Khim2020II} that the misalignments induced by secular evolution may include cases of outflowing gas driven by AGN feedback or star formation.

Overall, our results appear to be at odds with the findings of \cite{Khim2020II}. However, this discrepancy is likely to showcase, as previously noted, that a potentially significant fraction of our misaligned galaxies classified as being caused by gas accretion from other sources are the result of past mergers. The other notable difference is in the fraction of environmentally driven misalignments, although in this case the two classifications are not comparable at face value. It was also noted by \cite{Khim2020II} that some environmental driven misalignments in their sample show a temporary increase in their gas mass, contrary to what would be expected from ram-pressure stripping. 
Overall, this comparison highlights the difficulty in identifying the origins of stellar-gas misalignments in both observations and simulations, and how multiple physical processes are expected to play a role in shaping the properties of our kinematically decoupled sample.

\subsubsection{The dependency of misaligned fractions on stellar mass and SFR}
The results presented in Figure \ref{fig:Fraction_plots} show an increase in misalignment fractions with stellar mass between $10^9 - 10^{11}\ \rm{M_{\odot}}$ (top left), and a decrease with SFR between $0.5\lesssim$ log(SFR / $\rm{M_{\odot}\ yr^{-1}}$)$\ \lesssim -2$. Outside of these ranges, the significance of our results is affected by uncertainties pertaining to low number statistics. 

Previous studies of galaxies with kinematically decoupled stellar-gas rotation, both observational (e.g. \citealt{Davis2011}; \citealt{Jin2016}; \citealt{Zhou2022}) and simulation-based (\citealt{CdpLagos2015}) have reported increasing misalignment fractions with stellar mass up to a turnover value, followed by a steep decrease. 
\cite{Davis2011} mentioned that a potential explanation for low  fractions of kinematically decoupled objects at high $M_{\star}$ could be the presence of AGN and/or X-ray halos which prevent external gas from entering the galaxy, especially in the cold accretion mode. Our results do not recover a turnover in the increasing trend of misalignment fractions with $M_{\star}$ up to $10^{11}\ \rm{M_{\odot}}$, although we cannot exclude the possibility of a drop at higher stellar masses, given the number statistics available. 
The decrease in misalignment fractions with SFR for our sample is qualitatively consistent with that reported by \cite{Jin2016}, within the range where our results are statistically significant.

The observed trends remain the same when excluding all outflows or Seyfert objects from our sample, thus ruling out the possibility that AGN feedback is resulting in a decrease in misalignment fractions, at least up to $10^{11}\ \rm{M_{\odot}}$. This result is still in relative agreement with \cite{Davis2011}, who reported a decrease in misalignment fractions beginning around $M_{\rm{K}} \approx -24\ $mag ($\approx 10^{11}\ \rm{M_{\odot}}$; similarly for \citealt{Jin2016}), although this turnover was found to be at $\sim 10^{10.5}\ \rm{M_{\odot}}$ by \cite{CdpLagos2015} and \cite{Zhou2022}.
\section{Summary \& Conclusions}

In this paper, we computed stellar and gas kinematic PAs for a sample of 1445 galaxies in the nearby Universe selected from SAMI DR3. Following this, we identified a sample of 169 galaxies showing kinematic misalignments between the stellar and gas components. We compared the distributions of stellar-gas PA offset angles between misaligned sub-samples with different morphological and star-forming properties. We identified the main ionisation mechanisms in misaligned galaxies and used this information to determine the most probable physical process causing the kinematic decoupling. Finally, we analysed the degree to which $n_{\rm{S\Acute{e}rsic}}$, $\lambda_{Re}$ and sSFR correlate with the prevalence and timescales of misalignments. Our main results are as follows:

\begin{itemize}
    
    \item \textbf{Misalignment fractions increase towards larger $M_{\star}$ and lower SFR; the trends are unchanged when considering only accretion-driven cases.} The majority of misaligned galaxies are located below the SFMS, at high stellar masses and S\'ersic indices (Figure \ref{fig:Mstar_SFR_nsersic_plots}). The fractions of misaligned galaxies increases towards larger $M_{\star}$ in the range $10^{9-11}\ \rm{M_{\odot}}$, and decreases towards larger SFR, between $10^{-2}\ \rm{M_{\odot}\ yr^{-1}}$ and $10^{0.5}\ \rm{M_{\odot}\ yr^{-1}}$ (Figure \ref{fig:Fraction_plots}). 
    
    The trends observed did not change after excluding all outflows/Seyfert objects in our misaligned sample, showing that AGN feedback disturbing accreted gas is not resulting in lower observed misaligned fractions towards high stellar masses (up to ${10^{11}}\ \rm{M_{\odot}}$). 
    
    \item \textbf{Misalignment prevalence and timescales of gas settling are associated with both morphology and sSFR.} Statistical differences in the $|\Delta PA_{\rm{stars-gas}}|$ distributions between the late- and early-type misaligned sub-samples cannot be explained by morphology alone. The same differences
    are found between the SF and non-SF misaligned populations, indicating
    that viscous drag forces between pre-existing and newly acquired
    gas are also influencing this process. Misaligned galaxies have higher $n_{\rm{S\Acute{e}rsic}}$ and lower $\lambda_{Re}$ than aligned galaxies matched in $M_{\star}$ and SFR, as well as lower sSFR when controlling for $M_{\star}$ and $n_{\rm{S\Acute{e}rsic}}$. While the latter is valid for the early-type population separately, the same is not found for the late-type misaligned galaxies, with the caveat of low-number statistics. Overall, we found evidence that the prevalence and timescales of misalignments are significantly affected by an interplay between morphology and sSFR. If confirmed, our trends with sSFR for late-type galaxies could provide evidence for disc-like morphologies being the result of preferentially aligned gas accretion at higher redshifts.
    
    \item \textbf{The counter-rotating misaligned configuration is found to be stable.} We find a statistically significant counter-rotating peak in the position angle offset distribution around $150^{\rm{o}}-180^{\rm{o}}$ (Figure \ref{fig:dPA_plots}), providing observational evidence of the stability of this configuration. The relative size of the counter-rotating peak in comparison to the co-rotating one (4.2 \% for our whole misaligned sample) was found to be smaller than previously predicted by semi-analytic models of disc formation ($\sim$ 10 \%, \citealt{Stevens2016}). This difference can be explained by considering the effect of drag forces between newly acquired and pre-existing gas components, since this would also affect gas accreted in a counter-rotating orbit. 

    \item \textbf{Newly accreted gas is not feeding star formation in most (accretion-driven) misaligned galaxies.} We provided a spectral classification of misaligned galaxies based on the main gas ionisation mechanisms, into star-forming, LINERS and Seyferts (Figure \ref{BPT_plots}). The result of this classification is summarised in Table \ref{tab:tablemaster}. 
    When considering only accretion-driven misalignments, merely 24\% of such objects have star formation as the main gas ionisation mechanism, indicating that the newly acquired gas is not feeding new star formation in most (76\%) cases.

    \item \textbf{Stellar-gas kinematic misalignments in the nearby Universe are caused by both gas accretion (from filaments, outer halo, mergers or galaxy interactions) as well as gas outflows driven by AGN feedback.} In this work, we identified the most probable physical process causing the kinematic decoupling in our misaligned sample (Figure \ref{fig:Misalignment_cause_plot}). 
    We found that such features are caused by: outflows (8 \%); ongoing/recent gas-rich mergers (14 \%); tidal interactions (3 \%); gas accretion from any other sources, including past mergers (74 \%) and ram-pressure stripping (< 1 \%). 51 misaligned galaxies in our sample are group satellites (45 of which are not classified as outflows, mergers or tidally interacting), which sets an upper limit to the number of cases where environmental processes (e.g. gas stripping) could play a direct role. 
    
\end{itemize}

Our results have shown show that kinematic misalignments in the nearby Universe are caused by both gas accretion (from filaments, outer halo, mergers or interactions) and gas being expelled from the galaxy in an outflow. For accretion-driven cases, the new gas does not appear to be feeding star formation, in most cases.
The likelihood and timescales of misalignments are correlated with both morphology and sSFR, with misaligned galaxies typically having higher $n_{\rm{S\Acute{e}rsic}}$ as well as lower $\lambda_{Re}$ and sSFR than appropriately matched samples of aligned galaxies. Our results also show marginal evidence (given the low number statistics) in favor of the scenario in which late-type morphologies are produced by preferentially aligned gas accretion. 
The interpretation of our results would be facilitated by a better theoretical understanding of the expected contribution of morphology and gas content to driving misalignment timescales, potentially achieved with the aid of hydrodynamic simulations. 

\section*{Acknowledgements}

We thank the anonymous reviewer for a constructive report which improved the quality and impact of this work.

\textbf{AR} acknowledges that this research was carried out while the author was in receipt of a Scholarship for International Research Fees (SIRF) and an International Living Allowance Scholarship (Ad Hoc Postgraduate Scholarship) at The University of Western Australia. This research was conducted by the Australian Research Council Centre of Excellence for All Sky Astrophysics in 3 Dimensions (ASTRO 3D), through project number CE170100013.
\textbf{LC} acknowledges support from the Australian Research Council Discovery Project and Future Fellowship funding schemes (DP210100337, FT180100066). \textbf{SB} acknowledges funding support from the Australian Research Council through a Future Fellowship (FT140101166). \textbf{JJB} acknowledges support of an Australian Research Council Future Fellowship (FT180100231). \textbf{JBH} is supported by an ARC Laureate Fellowship FL140100278. The SAMI instrument was funded by Bland-Hawthorn's former Federation Fellowship FF0776384, an ARC LIEF grant LE130100198 (PI Bland-Hawthorn) and funding from the Anglo-Australian Observatory. \textbf{MSO} acknowledges the funding support from the Australian Research Council through a Future Fellowship (FT140100255). \textbf{JvdS}  acknowledges support of an Australian Research Council Discovery Early Career Research Award (project number DE200100461) funded by the Australian Government.

The SAMI Galaxy Survey is based on observations made at the Anglo-Australian Telescope. The Sydney-AAO Multi- object Integral-field (SAMI) spectrograph was developed jointly by the University of Sydney and the Australian Astronomical Observatory. The SAMI input catalogue is based on data taken from the Sloan Digital Sky Survey, the GAMA Survey and the VST ATLAS Survey. The SAMI Galaxy Survey is supported by the Australian Research Council Centre of Excellence for All Sky Astrophysics in 3 Dimensions (ASTRO 3D), through project number CE170100013, the Australian Research Council Centre of Excellence for All-sky Astrophysics (CAASTRO), through project number CE110001020, and other participating institutions. The SAMI Galaxy Survey website is http://sami-survey.org/.

\section*{Data availability}

The SAMI data cubes and value-added products used in this paper are available from Astronomical Optics’ Data Central service at: \url{https://datacentral.org.au/}. Computed stellar and gas PAs, stellar masses and SFRs can be provided upon request to the author. Physical properties of the misaligned galaxies identified in this study are outlined in Table \ref{tab:TableA2} (Appendix \ref{A4}).

\section*{Author contribution statement}

This project was devised by \textbf{AR}, \textbf{LC} and \textbf{AFM}. \textbf{AR} performed the analysis and drafted the paper. \textbf{JB}, \textbf{SC}, \textbf{SR}, \textbf{JvdS}, \textbf{JBH}, \textbf{MO}, and \textbf{JL} provided key support to all the activities of the SAMI Galaxy Survey (`builder status’). \textbf{LC}, \textbf{AFM}, \textbf{SB}, \textbf{JB}, \textbf{BC}, \textbf{SC}, \textbf{BG}, \textbf{SR}, \textbf{JvdS}, \textbf{JBH} and \textbf{MO} discussed the results and commented on the manuscript.


\bibliographystyle{mnras}
\bibliography{Bibliography.bib} 




\appendix
\section{Misalignment fraction statistics}
\label{A0}
A breakdown of the fractions of parent sample galaxies in different misalignment ranges, split between the SF/non-SF and late/early-type populations, is shown in Table \ref{tab:TableA1}. Uncertainties in the fractions reflect the number of galaxies which could become misaligned/aligned, assuming an average uncertainty of $\pm\ 9^{\rm{o}}$ in $|\Delta PA_{\rm{stars-gas}}|$. This estimate is computed as the average between the $68^{\rm{th}}$ percentiles of the |$PA_{\rm{Kin}}-PA_{\rm{Radon}}$| distribution for stars and gas rotation, i.e. $(9.0^{\rm{o}})_{\rm{stars}}$ and $(9.5^{\rm{o}})_{\rm{gas}}$, excluding the galaxies with discrepancies larger than 1 $\sigma$ in the $PA_{\rm{Kin}}-PA_{\rm{Radon}}$ distributions for either stars or gas (see Section \ref{sec:PAs_measurements}) as well as the kinematically unconstrained galaxies in our misaligned sample (see Section \ref{sec:misaligned}). The table also shows the ratios between the sizes of the counter- and co- rotating peaks in the $|\Delta PA_{\rm{stars-gas}}|$ distributions, for each respective galaxy population.

\setlength{\tabcolsep}{4.4pt} 
\renewcommand{\arraystretch}{1.1}
\begin{table*}
\centering
\caption{Fractions of parent sample galaxies (split into star-forming/non star-forming and late/early-types) which are in a given misalignment category (as shown by the misalignment angular range in brackets), and the counter-rotating to aligned galaxy number ratio. The `\textit{All misaligned}` category includes the kinematically unconstrained misaligned galaxies (see Section \ref{sec:misaligned}) }\label{tab:TableA1}
\begin{tabular*}{\textwidth}{@{\extracolsep{\fill}}cccccc} 
\hline\multirow{2}{*}{Misalignment range} & \multicolumn{5}{c}{Fraction of parent (sub)sample galaxies that are in a given misalignment range} \\
                             & Whole          & SF          & Non SF          & Late-type          & Early-type         \\
\hline
All misaligned [$30^{\rm{o}}-180^{\rm{o}}$] (+ kin. unc.)              &       $11.7\substack{+2.7 \\ -0.9}$\% (169/1445)         &    $4.4\substack{+2.3 \\ -0.6}$\% (43/982)          &   $31\substack{+4 \\ -2}$\% (120/387)               &     $3.0\substack{+2.4 \\ -0.3}$\% (28/947)                &       $30\substack{+3 \\ -2}$\% (137/452)              \\
Unstable [$30^{\rm{o}}-150^{\rm{o}}$]                     & $6.6\substack{+3.1 \\ -0.9}$\% (95/1445)                &   $1.7\substack{+2.5 \\ -0.4}$\% (17/982)           &    $20\substack{+5 \\ -2}$\% (76/387)              &       $1.0\substack{+2.0 \\ -0.2}$\% (9/947)              &        $19\substack{+4 \\ -2}$\% (84/452)                  \\
Stable/counter-rotating ($150^{\rm{o}}-180^{\rm{o}}$]     & $3.7\substack{+0.1 \\ -0.4}$\% (53/1445) &           $1.9\substack{+0.1 \\ -0.2}$\% (19/982)    &        $8.3\substack{+0.3 \\ -1.0}$\% (32/387)     &       $1.4\substack{+0.1 \\ -0.1}$\% (13/947)          &      $8.4\substack{+0.2 \\ -1.1}$\% (38/452)                        \\         

Aligned [$0^{\rm{o}}-30^{\rm{o}}$)               &       $88.3\substack{+0.8 \\ -2.7}$\% (1276/1445)         &    $95.6\substack{+0.6 \\ -2.3}$\% (939/982)          &   $69\substack{+2 \\ -4}$\% (267/387)               &     $97.0\substack{+0.3 \\ -2.4}$\% (919/947)                &       $70\substack{+2 \\ -3}$\% (315/452)              \\
\hline
Counter-rotating to aligned ratio              &       4.2$\substack{+0.2 \\ -0.5}$\%        &    2.0$\substack{+0.1 \\ -0.3}$\%         &   12$\substack{+1 \\ -2}$\%             &     1.4$\substack{+0.2 \\ -0.1}$ \%                 &       12.1$\substack{+0.7 \\ -2.0}$ \%            \\

\hline

\end{tabular*}
\end{table*}


\section{Galaxy properties driving misalignment timescales}
\label{A1}

\begin{figure}
\centering	\includegraphics[width=6.9cm]{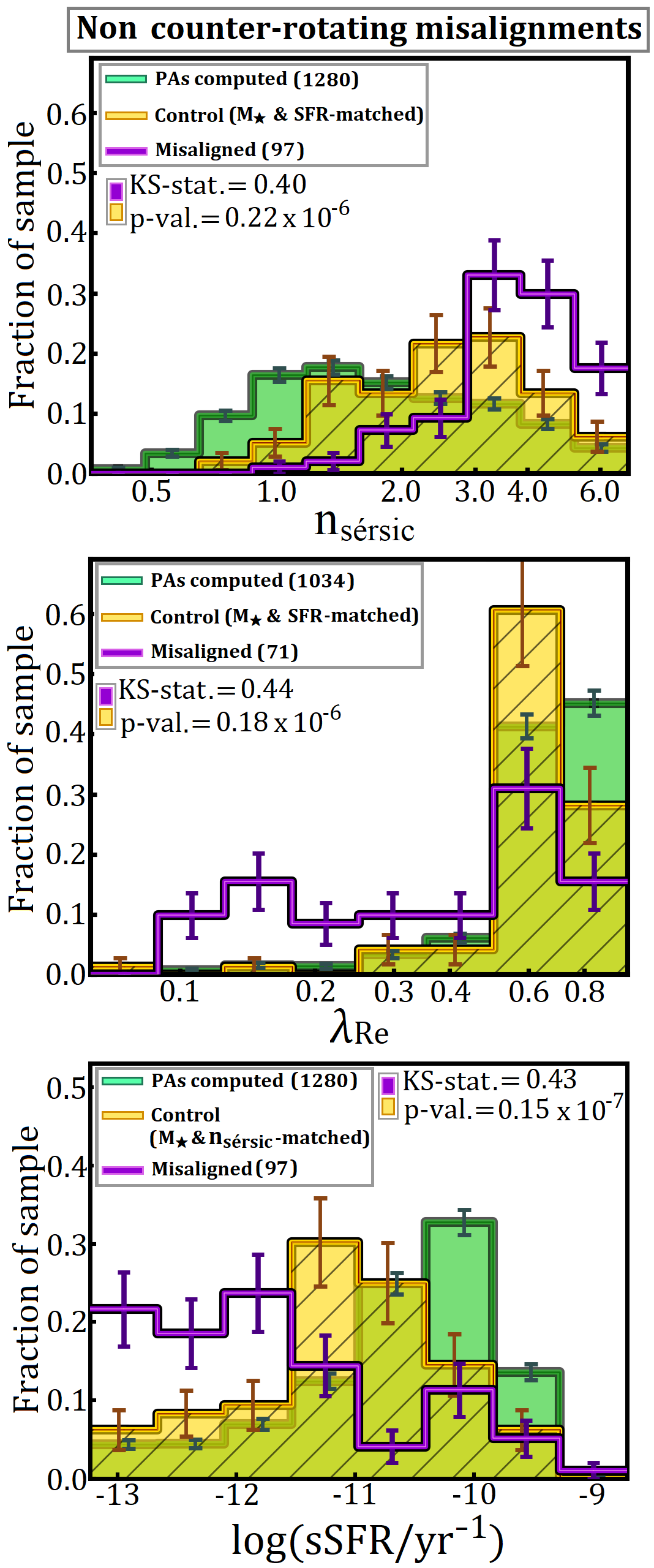}    \caption{(\textbf{Top \& Middle}) Same as Figure \ref{fig:nsersic_lambda_re_dist} - top row, but excluding the counter-rotating galaxies from the misaligned and parent samples. (\textbf{Bottom}) Same as Figure \ref{fig:sSFR_Dist} - top, but again excluding the counter-rotating galaxies from the misaligned and parent samples. Note that fractions on the vertical axis are calculated with respect to the revised misaligned and parent samples (i.e. excluding the counter-rotating misalignments).}
    \label{fig:A1}
\end{figure}

The results presented in Section \ref{sec:distributions_results} include our (non-outflow) misalignments across the whole range of kinematic PA offsets. Galaxies with $|\Delta PA_{\rm{stars-gas}}|$ between $30^{\rm{o}}-150^{\rm{o}}$ are considered to involve gas still in the process of aligning with the stellar component, while counter-rotating systems ($150^{\rm{o}}-180^{\rm{o}}$)] have already reached a stable state, within uncertainties (or alternatively, gas was initially accreted in a retrograde orbit). 

We split our misaligned sample into stable (counter-rotating) and unstable (non counter-rotating) configurations and re-plot the distributions in $n_{\rm{S\Acute{e}rsic}}$, $\lambda_{Re}$ and sSFR for the latter in Figure \ref{fig:A1}. This approach is expected to show how the plotted galaxy properties are influencing gas settling timescales.
The trends are statistically similar to those in Figures \ref{fig:nsersic_lambda_re_dist} (top row) and \ref{fig:sSFR_Dist} (top), confirming that these are driven by misaligned galaxies in unstable configurations (non counter-rotating), rather than by counter-rotating objects. This finding shows that misalignment timescales are longer in galaxies with early-type morphologies (higher $n_{\rm{S\Acute{e}rsic}}$ and lower $\lambda_{Re}$) and lower sSFR.

\section{Misalignment angle distributions without outflow-driven cases and Seyfert objects}
\label{A2}

\begin{figure}
\centering	\includegraphics[width=\columnwidth]{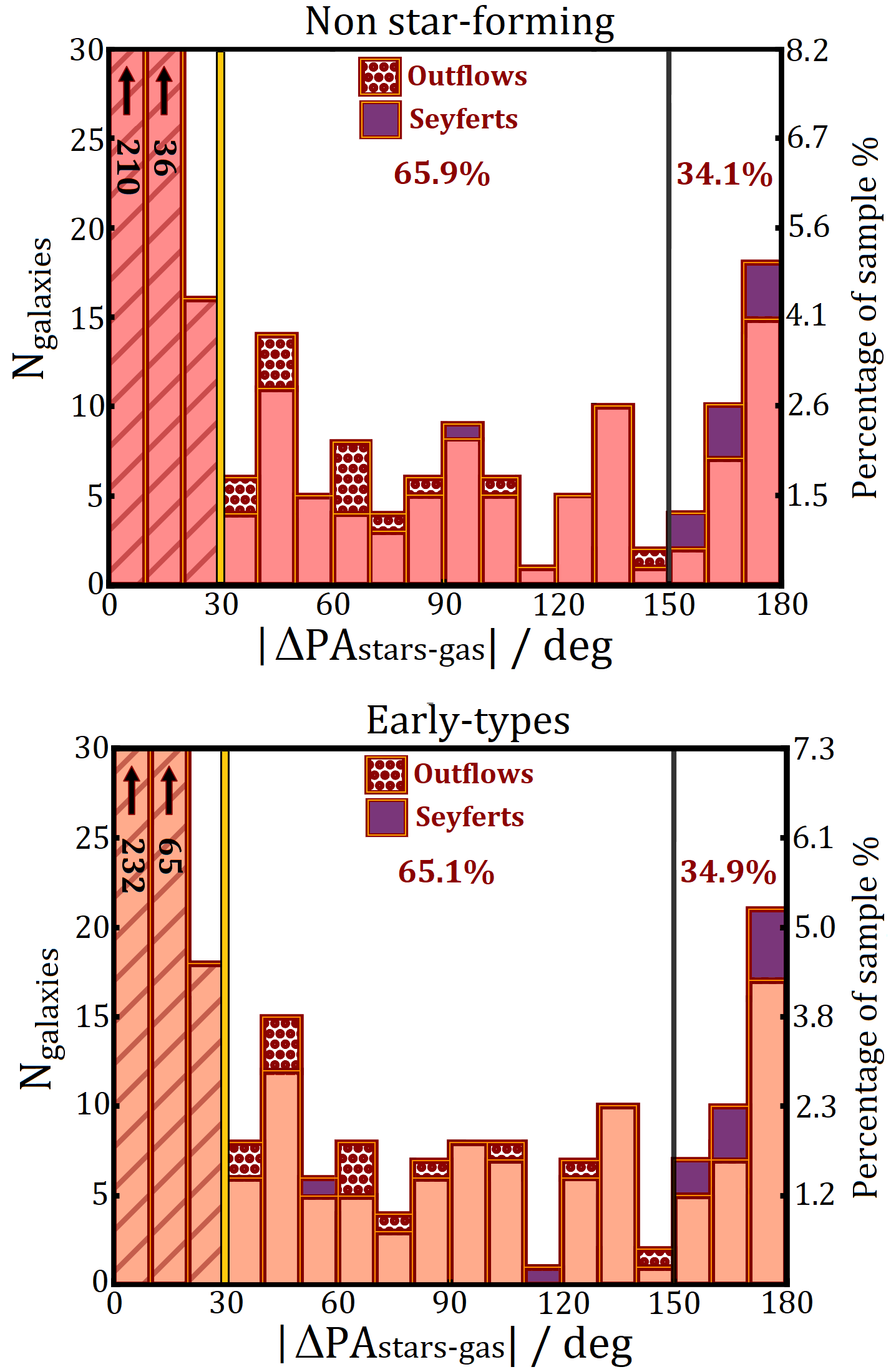}  \caption{Same distributions as in the right column of Figure \ref{fig:dPA_plots}, highlighting the contribution of Seyfert objects and outflows to observed misaligned numbers. The percentages displayed on the plots show the fraction of early-type misaligned objects in each category after excluding all outflow-driven cases.}
\label{fig:A2}
\end{figure}

Figure \ref{fig:A2} shows the contribution of outflows and Seyfert objects to observed misalignments in the non-SF and early-type sub-samples. We found that the exclusion of such galaxies does not change the statistical significance of the results in Section \ref{sec:dpa_distribution}. The three Seyfert objects outside of the counter-rotating region, which are not classified as outflows are: \textbf{CATID:618993}, which displays clear signs of having recently suffered a head-on collision with another galaxy, which likely shock-heated the gas in the central parts, resulting in the observed spectral properties (this object is classified as SF and does not appear on the top panel); \textbf{CATID:9008500100}, which was identified as having a misalignment due to gas stripping in the cluster environment (this object does not have a SFR estimate and does not appear on the top panel); \textbf{CATID:56181}, which is identified as having an uncertain spectral classification (Section \ref{sec:spectral_results}; this object is classified as late-type and does not appear on the bottom panel).

\section{Properties of misaligned galaxies}
\label{A4}

The properties of misaligned galaxies in our sample are outlined in Table \ref{tab:TableA2}: Right ascension (RA); declination (DEC); stellar ($PA_{\rm{stars}}$) and gas ($PA_{\rm{gas}}$) kinematic position angles (Section \ref{sec:PAs_measurements}); log of SFR and stellar mass (Section \ref{sec:Ms_SFR}); the most probable physical cause of the kinematic misalignment, as identified in Section \ref{sec:misalign_cause}.

\setlength{\tabcolsep}{9.3pt} 
\renewcommand{\arraystretch}{1.2}
\begin{table*}
\centering
\caption{Summary of the properties of misaligned galaxies in our sample. Objects without a recorded stellar and/or gas PA have unconstrained kinematics in the respective component. }\label{tab:TableA2}

\begin{tabular*}{\linewidth}{cccccccc}
\hline
\multicolumn{1}{l}{CATID} & \multicolumn{1}{l}{RA [deg]} & \multicolumn{1}{l}{DEC [deg]} & \multicolumn{1}{l}{$PA_{\rm{stars}}$ [deg]} & \multicolumn{1}{l}{$PA_{\rm{gas}}$ [deg]} & \multicolumn{1}{l}{log(SFR [$\rm{M_{\odot}yr^{-1}}$])} & \multicolumn{1}{l}{log($M_{\star}$ [$\rm{M_{\odot}}$])} & \multicolumn{1}{l}{Misalignment cause} \\
\hline
7139                      & 175.8283              & +0.7479                & 86                            & 3                           & -1.48
& 10.62
& Accretion                   \\
7715                      & 179.2287              & +0.7239                 & 183                           & 348                         & -0.99
& 11.08
& Merger                    \\
7839                      & 179.6154              & +0.7179                & 232                           & 295                         & -0.93
& 11.18
& Outflow                    \\
7992                      & 180.2212              & +0.8138                & 208                           & 336                         & -0.43
& 11.03
& Accretion                    \\
8487                      & 182.6359              & +0.6775                & 271                           & 17                          & -0.93
& 10.71
& Accretion                   \\ 
8865                      & 184.0313              & +0.6563                & -                           & 328                         & -1.14
& 10.72
& Accretion                    \\
9062                      & 184.8551             & +0.6310                & 168                           & 101                         & -1.62
& 10.31
& Accretion                    \\
9352                      & 185.9772             & +0.8305                 & 7                             & 208                         & -0.63
& 8.99               & Accretion                   \\
17390                     & 223.3072              & +0.6565                & 17                            & 245                         & -0.61
& 11.21
& Merger                    \\
22633                     & 178.4448              & +1.1935                & 156                           & 118                         & 0.81
& 10.51
& Tidal interaction                    \\
23265                     & 180.9158               & +1.2046                & 277                           & 346                         & -1.78
& 11.05
& Accretion                   \\
30847                     & 177.1550             & -1.2128               & 122                           & 59                          & -1.47
& 11.65
& Outflow                   \\
31740                     & 181.3128              & -1.2196              & 39                            & 1                           & 0.45
& 9.87
& Accretion                    \\
39057                     & 175.0267               & -0.8377             & 103                           & 298                         & -2.32
& 10.53
&  Accretion                   \\
39365                     & 176.6795               & -0.6783              & 123                           & 187                         & -0.85
& 9.24
& Accretion                    \\
41059                     & 184.2046               & -0.7418               & -                           & 340                         & -1.50
& 11.43
& Accretion                    \\
41274                     & 185.1205              & -0.6882               & 120                           & 156                         & -2.31
& 10.70
& Accretion                   \\
47493                     & 213.1036             & -0.8320               & 239                           & 271                         & 0.28
& 9.79
&  Accretion                   \\
48049                     & 215.6319              & -0.6553               & 75                            & 244                         & -0.57
& 10.80
& Accretion                    \\
49784                     & 222.5731              & -0.6501               & 299                           & 91                          & -2.66
& 10.35
& Accretion                   \\
54918                     & 179.9836              & -0.3231               & 0                             & 314                         & -1.69
& 11.31
& Tidal interaction                   \\
55143                     & 180.6230              & -0.2124               & 17                            & 305                         & -1.81
& 11.20
&   Accretion                 \\
55245                     & 181.0795             & -0.3155               & -                            & -                         & 0.32
& 11.20
&  Merger                  \\
56055                     & 184.3767              & -0.2320               & 91                            & 156                         & -0.45
& 9.36
& Accretion                   \\
56181                     & 184.8369              & -0.2415               & 238                           & 331                         & -2.57
& 10.43
& Accretion                    \\
65410                     & 223.0809               & -0.2692               & 80                            & 247                         & -0.17
& 10.21
& Accretion                    \\
69740                     & 176.3596              & +0.0040                & 28                            & 349                         & 0.06
& 9.68
& Accretion                    \\
70114                     & 178.5511              & +0.1366                & 21                            & -                          & -0.76
& 9.11
& Merger                   \\
70532                     & 180.2708               & +0.1138                & 317                           & 128                         & -1.32
& 10.31
& Accretion                   \\
70670                     & 180.8682               & +0.0601                & -                           & 343                         & -1.74
& 11.35
& Accretion                    \\
79810                     & 223.2611              & +0.2031                & 116                           & 197                         & -0.06
& 9.88
&  Accretion                   \\
84680                     & 179.3750             & +0.6026                & 275                           & 91                          & -1.91
& 10.58
& Accretion                    \\
85205                     & 181.4555             & +0.5402                & -                           & 195                         & -0.85
& 10.25
&  Accretion                   \\
91545                     & 212.6103              & +0.4441                & 216                           & 317                         & -1.32
& 10.53
&  Accretion                  \\
91568                     & 212.7591              & +0.6085                & 181                           & 132                         & -1.55
& 10.00
& Outflow                   \\
91691                     & 212.9820             & +0.4666                & 83                            & 40                          & -1.29
& 10.74
&  Accretion                  \\
91697                     & 213.0902              & +0.5304                & 210                           & 24                          & -1.83
& 10.72
& Accretion                   \\
91996                     & 214.4757              & +0.4614                & 103                           & 280                         & -1.64
& 10.67
& Accretion                    \\
92592                     & 216.9740              & +0.5696                & 79                            & 307                         & -1.72
& 11.14
& Merger                    \\
93807                     & 222.5090              & +0.5788                 & 1                             & 179                         & -0.64
& 10.68
& Accretion     \\             
105519                     & 212.3401             & +1.0277                & 222                             & 86                         & -2.67
& 10.43
& Accretion     \\ 

106549                     & 216.5541             & +0.8606                & 256                             & 55                         & 0.13
& 10.31
& Accretion     \\ 

107137                     & 218.6906            & +0.8887                & 218                             & -                         & -1.55
& 10.21
& Accretion     \\

136308                     & 174.4956            & -1.6946                & 41                             & 229                         & -1.14
& 10.63
& Accretion     \\ 

136616                     & 175.9020           & -1.6383                & 106                             & 332                         & -1.82
& 10.74
& Accretion     \\

136842                    & 176.2912               & -1.7626                 & 67                            & 235                         & -1.04
& 10.85
& Merger             \\
136880                    & 176.3371               & -1.8128                 & 71                            & 324                         & -2.34
& 9.86
& Accretion          \\
143227                    & 174.8290               & -1.3384                 & 265                           & 91                          & -0.64
& 10.67
& Accretion          \\
144320                    & 179.7335               & -1.4304                 & 172                           & 356                         & 0.20
& 10.34
& Accretion          \\
144335                    & 179.7592               & -1.3131                 & -                           & 93                          & -1.25
& 11.31
& Accretion          \\
184202                    & 175.4762               & -1.4559                 & 36                            & 0                           & -1.44
& 10.98
& Outflow            \\
184648                    & 177.4014               & -1.4555                 & 241                           & 62                          & -1.91
& 10.66
& Merger             \\
202399                    & 129.4574               & -0.2516                 & 307                           & 262                         & -1.93
& 11.07
& Accretion         

\end{tabular*}
\end{table*}

\renewcommand\tablename{C1}

\renewcommand\thetable{{\ref{tab:TableA2}}}
\setlength{\tabcolsep}{9.3pt} 
\renewcommand{\arraystretch}{1.2}
\begin{table*}
\centering
\caption{(Continued)}\label{tab:TableA3}
\setcounter{table}{30}
\begin{tabular*}{\linewidth}{cccccccc}
\hline
\multicolumn{1}{l}{CATID} & \multicolumn{1}{l}{RA [deg]} & \multicolumn{1}{l}{DEC [deg]} & \multicolumn{1}{l}{$PA_{\rm{stars}}$ [deg]} & \multicolumn{1}{l}{$PA_{\rm{gas}}$ [deg]} & \multicolumn{1}{l}{log(SFR [$\rm{M_{\odot}yr^{-1}}$])} & \multicolumn{1}{l}{log($M_{\star}$ [$\rm{M_{\odot}}$])} & \multicolumn{1}{l}{Misalignment cause} \\
\hline

208652                    & 129.7534 & +0.0623 & 71                            & 118                         & -1.88
& 11.12
& Accretion          \\
209680                    & 134.4890 & +0.1712 & -                           & 227                         & -1.76
& 11.16
& Accretion          \\
210375                    & 138.2259 & +0.0908 & 29                            & 159                         & -1.26
& 10.59
& Accretion          \\
210611                    & 138.9173 & +0.1856 & 260                           & 68                          & 0.27
& 10.63
& Accretion          \\
216184                    & 137.1743 & +0.4847 & 36                            & 146                         & -1.33
& 10.56
& Accretion          \\
218713                    & 140.0100 & +0.8442 & 46                            & 241                         & 0.03
& 10.09
& Accretion          \\
220372                    & 181.2894 & +1.5593 & -                           & 180                         & -0.83
& 9.02
& Accretion          \\
220499                    & 181.8585 & +1.5456 & 86                            & 188                         & -0.94
& 10.33
& Outflow            \\
227266                    & 212.8593 & +1.2865 & 48                            & 176                         & -0.57
& 11.05
& Outflow            \\
227278                    & 212.8299 & +1.3095 & 109                           & 178                         & -2.12
& 10.20
& Accretion          \\
227617                    & 214.3240 & +1.2215 & 115                           & 63                          & -1.11
& 10.59
& Accretion          \\
227905                    & 215.3194 & +1.2074 & 80                            & 36                          & -1.47
& 10.88
& Accretion          \\
228564                    & 217.8525 & +1.2344 & 178                           & 72                          & -1.48
& 10.69
& Accretion          \\
229220                    & 221.2286 & +1.0398 & 29                            & 71                          & -3.54
& 9.46
& Outflow            \\
230421                    & 179.7718 & +1.9099 & 89                            & 271                         & -0.17
& 10.73
& Accretion          \\
230556                    & 180.3270 & +1.8928 & -                           & 139                         & -0.82
& 11.17
& Accretion          \\
231528                    & 184.2065 & +1.9079 & 271                           & 133                         & -0.89
& 10.90
& Merger             \\
238453                    & 214.3032 & +1.7140 & 230                           & 187                         & -0.59
& 10.88
& Accretion          \\
238922                    & 215.7873 & +1.6716 & 132                           & 229                         & -1.39
& 10.76
& Merger             \\
272820                    & 180.9651 & +1.2951 & 0                             & 157                         & -0.20
& 11.03
& Accretion          \\
272822                    & 180.9982 & +1.4109 & 156                           & 115                         & 0.09
& 10.17
& Merger             \\
272831                    & 181.0082 & +1.4449 & 286                           & 48                          & -0.31
& 11.22
& Merger             \\
278840                    & 133.9595 & +0.8311 & 40                            & 226                         & 0.01
& 10.52
& Accretion          \\
279905                    & 139.9800 & +0.9451 & 173                           & 298                         & -1.25
& 10.51
& Accretion          \\
290071                    & 184.5093 & +1.7951 & -                           & 318                         & -0.52
& 9.40
& Accretion          \\
296725                    & 213.0289 & +1.5178 & 56                            & 226                         & -0.25
& 9.71
& Accretion          \\
296927                    & 214.0178 & +1.4582 & 151                           & 15                          & -1.24
& 11.07
& Merger             \\
298590                    & 220.6776 & +1.3198 & 68                            & 307                         & -0.08
& 10.59
& Accretion          \\
300406                    & 129.4051 & +1.0731 & 284                           & 52                          & -1.96
& 11.04
& Accretion          \\
300411                    & 129.4859 & +1.0724 & -                           & -                         & -2.07
& 10.93
& Accretion          \\
300691                    & 130.4123 & +1.1166 & 197                           & 117                         & -2.54
& 10.46
& Outflow             \\
300787                    & 130.9350 & +1.0792 & 301                           & 128                         & 0.08
& 10.43
& Accretion          \\
301381                    & 133.6175 & +1.2020 & 55                            & 24                          & 0.77
& 10.70
& Tidal interaction  \\
301937                    & 135.6532 & +1.3202 & -                           & -                         & 0.06
& 11.06
& Merger             \\
302684                    & 138.8110 & +1.5107 & 13                            & 201                         & -0.04
& 10.47
& Accretion          \\
319018                    & 213.3043 & +1.7828 & 168                           & 133                         & 0.11
& 10.16
& Accretion          \\
319293                    & 214.0075 & +1.9166 & 47                            & 185                         & -2.62
& 10.37
& Accretion          \\
321059                    & 220.9849 & +1.6042 & 60                            & 252                         & -1.85
& 9.77
& Accretion          \\
323593                    & 132.3879 & +1.7280 & 115                           & 7                           & -0.12
& 11.18
& Merger             \\
324351                    & 135.7165 & +1.7110 & 119                           & 298                         & -0.64
& 10.56
& Accretion          \\
347471                    & 137.8833 & +2.2398 & 185                           & 260                         & -1.66
& 10.99
& Accretion          \\
371172                    & 130.4963 & +1.0173 & 17                            & 336                         & -1.61
& 11.27
& Accretion          \\
372408                    & 135.7051 & +1.0206 & 213                           & 36                          & -0.82
& 10.80
& Accretion          \\
380697                    & 129.5859 & +1.7325 & 183                           & 138                         & -0.43
& 11.11
& Accretion          \\
381206                    & 131.6918 & +1.8308 & -                           & 156                         & -0.59
& 11.26
& Accretion          \\
382154                    & 135.4670 & +1.9933 & 245                           & 143                         & -1.27
& 10.86
& Accretion          \\
386722                    & 133.6602 & +2.3210 & 202                           & 26                          & -0.03
& 10.44
& Accretion          \\
388558                    & 140.5922 & +2.5859 & 151                           & 60                          & -0.65
& 11.16
& Merger             \\
417440                    & 131.1279 & +2.2949 & 266                           & 3                           & -1.44
& 11.29
& Accretion          \\
422389                    & 130.6872 & +2.5831 & 279                           & 223                         & -1.69
& 10.64
& Accretion          \\
422649                    & 131.8277 & +2.7623 & 317                           & 122                         & -1.68
& 10.71
& Accretion          \\
460767                    & 213.7961 & -1.6352 & 34                            & 119                         & -1.32
& 10.27
& Accretion          \\
485833                    & 217.5687 & -1.7570 & 17                            & 54                          & 0.63
& 10.44
& Accretion         

\end{tabular*}
\end{table*}


\setlength{\tabcolsep}{9.0pt} 
\renewcommand{\arraystretch}{1.2}
\begin{table*}
\centering
\caption{(Continued)}\label{tab:TableA4}

\begin{tabular*}{\linewidth}{cccccccc}
\hline
\multicolumn{1}{l}{CATID} & \multicolumn{1}{l}{RA [deg]} & \multicolumn{1}{l}{DEC [deg]} & \multicolumn{1}{l}{$PA_{\rm{stars}}$ [deg]} & \multicolumn{1}{l}{$PA_{\rm{gas}}$ [deg]} & \multicolumn{1}{l}{log(SFR [$\rm{M_{\odot}yr^{-1}}$])} & \multicolumn{1}{l}{log($M_{\star}$ [$\rm{M_{\odot}}$])} & \multicolumn{1}{l}{Misalignment cause} \\
\hline

492486                    & 216.8988 & -1.3726  & 200                           & 160                         & -1.75
& 10.20
& Outflow            \\
492941                    & 218.8567 & -1.3991  & 7                             & 293                         & -1.38
& 10.44
& Outflow            \\
493838                    & 222.6039 & -1.1947  & 211                           & 129                         & -1.91
& 10.52
& Accretion          \\
505341                    & 219.3785 & -1.9642  & 312                           & 358                         & -0.60
& 10.78
& Accretion          \\
505765                    & 221.5508 & -1.8146  & 99                            & 7                           & -1.53
& 10.96
& Tidal interaction  \\
506119                    & 223.2724 & -1.7693  & 68                            & 255                         & -1.42
& 11.19
& Accretion          \\
508312                    & 216.5988 & -1.5937  & 132                           & 323                         & -0.85
& 10.35
& Accretion          \\
508480                    & 217.1822 & -1.6584  & 135                           & 331                         & -0.96
& 9.67
& Accretion          \\
508751                    & 218.1565 & -1.5301  & 54                            & 195                         & -1.12
& 10.34
& Outflow            \\
511863                    & 216.2875 & -1.1135  & 241                           & 71                          & -3.55
& 9.45
& Accretion          \\
514022                    & 214.0467 & -1.2685  & 315                           & 86                          & -1.56
& 10.25
& Accretion          \\
517278                    & 131.6832 & +2.5374  & 257                           & 312                         & -0.45
& 11.08
& Merger             \\
517279                    & 131.6804 & +2.5392  & 271                           & 7                           & -2.53
& 10.47
& Merger             \\
518829                    & 138.2295 & +2.9309  & 326                           & 130                         & -2.97
& 10.03
& Accretion          \\
534655                    & 174.3644 & -0.8466  & 331                           & 255                         & -0.44
& 11.21
& Accretion          \\
534904                    & 175.6776 & -0.9567  & 109                           & 296                         & -1.10
& 10.36
& Accretion          \\
536963                    & 183.5387 & -0.9386  & 183                           & 98                          & -1.96
& 11.04
& Accretion          \\
536994                    & 183.6069 & -0.9477  & 102                           & 339                         & 0.06
& 11.41
& Merger             \\
543537                    & 211.9120 & -0.9553  & 346                           & 67                          & -1.25
& 11.53
& Merger             \\
543895                    & 213.2723 & -0.8462  & 36                            & 224                         & 0.15
& 9.90
& Accretion          \\
549296                    & 131.0714 & -0.4249  & 277                           & 321                         & -1.47
& 10.96
& Accretion          \\
550781                    & 137.4683 & -0.5046  & 231                           & 46                          & -0.02
& 10.52
& Accretion          \\
551505                    & 140.5722 & -0.5469  & 11                            & 312                         & -0.44
& 11.49
& Accretion          \\
560208                    & 179.6707 & -0.5972  & 115                           & 73                          & -0.75
& 9.30
& Accretion          \\
560238                    & 179.8037 & -0.5238  & 208                           & 28                          & -0.13
& 10.90
& Accretion          \\
561488                    & 183.8700 & -0.4996  & 6                             & 113                         & -0.53
& 9.37
& Accretion          \\
570206                    & 222.7625 & -0.5271  & 3                             & 273                         & -0.56
& 10.70
& Accretion          \\
570240                    & 222.9383 & -0.5125  & 3                             & 143                         & -0.66
& 10.08
& Accretion          \\
573617                    & 129.2772 & -0.1871  & 91                            & 295                         & -0.22
& 10.18
& Accretion          \\
575339                    & 139.4849 & -0.1014  & 140                           & 3                           & -1.00
& 10.47
& Accretion          \\
585196                    & 181.6219 & -0.1724  & 336                           & 158                         & -1.31
& 10.68
& Accretion          \\
586330                    & 184.5455 & -0.1637  & 138                           & 340                         & -1.28
& 10.21
& Accretion          \\
592419                    & 212.9650 & -0.0509  & 55                            & 235                         & -0.67
& 10.18
& Accretion          \\
593642                    & 217.3625 & -0.2089  & 180                           & 3                           & -2.78
& 9.78
& Accretion          \\
594049                    & 218.9356 & -0.0134  & 197                           & 264                         & -1.81
& 11.20
& Outflow            \\
599834                    & 132.7113 & +0.3733  & 222                           & 128                         & -1.45
& 11.25
& Accretion          \\
599838                    & 132.7908 & +0.2536  & 221                           & 259                         & -0.71
& 11.08
& Outflow            \\
601105                    & 138.6037 & +0.3630  & 293                           & 136                         & -0.85
& 10.62
& Accretion          \\
609358                    & 175.4568 & +0.2160  & 156                           & 66                          & -0.50
& 11.02
& Merger             \\
609396                    & 175.5509 & +0.3345  & -                             & 97                          & 0.81
& 10.01
& Merger             \\
618151                    & 214.5170 & +0.2738  & 137                           & 314                         & -0.88
& 10.56
& Accretion          \\
618993                    & 217.6918 & +0.2477  & 7                             & 65                          & 0.68
& 11.03
& Tidal interaction  \\
623017                    & 136.3799 & +0.8264  & 175                           & 212                         & -1.67
& 10.98
& Accretion          \\
3634556                   & 139.4468 & -0.9859  & 216                           & 11                          & -0.31
& 11.12
& Accretion          \\
3913888                   & 138.1609 & -0.8696  & 17                            & 308                         & -1.23
& 10.43
& Outflow            \\
9008500001                & 10.4602  & -9.3032  & -                             & -                           & - & 11.87
& Accretion          \\
9008500100                & 10.3664  & -9.2298  & 102                           & 221                         & - & 10.34
& Gas stripping      \\
9008500326                & 10.4889  & -9.5899  & 61                            & 31                          & -0.90
& 10.31
& Accretion          \\
9008500391                & 10.1759  & -9.4781  & 43                            & 11                          & -1.43
& 10.73
& Accretion          \\
9008500848                & 10.7161  & -9.8078  & 95                            & 147                         & -1.27
& 10.78
& Accretion          \\
9011900101                & 14.1026  & -1.1429  & 46                            & -                           & -0.07
& 10.69
& Accretion          \\
9011900430                & 14.3954  & -1.3910  & -                             & 60                          & - & 11.52
& Accretion          \\
9091700033                & 355.371  & -29.1537 & 120                           & 296                         & - & 10.84 & Accretion

\end{tabular*}
\end{table*}


\setlength{\tabcolsep}{9.0pt} 
\renewcommand{\arraystretch}{1.2}
\begin{table*}
\centering
\caption{(Continued)}\label{tab:TableA5}

\begin{tabular*}{\linewidth}{cccccccc}
\hline
\multicolumn{1}{l}{CATID} & \multicolumn{1}{l}{RA [deg]} & \multicolumn{1}{l}{DEC [deg]} & \multicolumn{1}{l}{$PA_{\rm{stars}}$ [deg]} & \multicolumn{1}{l}{$PA_{\rm{gas}}$ [deg]} & \multicolumn{1}{l}{log(SFR [$\rm{M_{\odot}yr^{-1}}$])} & \multicolumn{1}{l}{log($M_{\star}$ [$\rm{M_{\odot}}$])} & \multicolumn{1}{l}{Misalignment cause} \\
\hline

9091700469                & 355.9589               & -29.1907                & 200                           & 160                         & -0.42
& 10.00 
& Accretion          \\
9239900205                & 329.3300               & -8.0128                 & -                             & 293                         & -1.17
& 11.06
& Merger             \\
9239900315                & 329.1265               & -7.6720                 & 211                           & 129                         & 0.03
& 10.19
& Accretion          \\
9239900362                & 329.1797               & -7.5500                 & -                             & 358                         & -1.04
& 10.93
& Accretion          \\
9239900540                & 329.2880               & -7.3683                 & 99                            & 7                           & -2.50
& 10.47
& Accretion          \\
9239900555                & 329.8381               & -7.8469                 & 68                            & 255                         & -0.34
& 11.04
& Merger             \\
9239900821                & 329.9187               & -7.5172                 & 132                           & 323                         & -1.36
& 10.31
& Accretion          \\
9239900970                & 329.9294               & -7.3935                 & 135                           & 331                         & -1.17
& 9.97
& Accretion          \\
9239901001                & 329.9393               & -7.3867                 & 54                            & 195                         & -                       & 11.17
& Merger             \\
9388000135                & 337.2251               & -30.6287                & 241                           & 71                          & -                       & 11.73
& Accretion         

\end{tabular*}
\end{table*}


\bsp	
\label{lastpage}
\end{document}